\documentclass[aps,pra,floatfix,tightenlines,superscriptaddress,amsmath,amssymb]{revtex4}
\usepackage{graphicx}

\usepackage{bm}
\usepackage{color}
\definecolor{fgreen}{rgb}{0,0.7,0}
\definecolor{purple}{rgb}{1,0,1}

\usepackage{hyperref}
\makeatletter
\renewcommand*\env@matrix[1][*\c@MaxMatrixCols c]{%
  \hskip -\arraycolsep
  \let\@ifnextchar\new@ifnextchar
  \array{#1}}
\makeatother

\newcommand{\ket}[2] {| #1 \rangle_{#2}}
\newcommand{\bra}[2] {\langle #1 |_{#2}}
\newcommand{\ham}{\hat H}

\newcommand{\ee}[1] {\mathrm{e}^{#1}}
\newcommand{\hc}{\mathrm{h.c.}}
\newcommand{\cc}{\mathrm{c.c.}}
\definecolor{pink}{rgb}{1,0.078,0.57}

\newcommand{\dg}{^{\dagger}}

\begin{document}

\keywords{Two-dimensional spectroscopy, quantum optics, nonlinear optical spectroscopy.}


\title{Two-dimensional electronic spectroscopy for the quantum-optics enthusiast}
\author{Agata M. Bra\'nczyk$^1$, Daniel B. Turner$^1$ and Gregory D. Scholes}
\address{Department of Chemistry and Centre for Quantum Information and Quantum Control, 80 St. George Street, University of Toronto, Toronto, Ontario M5S 3H6 Canada}
%
%
  
\begin{abstract}
Recent interest in the role of quantum mechanics in the primary events of photosynthetic energy transfer has led to a convergence of nonlinear optical spectroscopy and quantum optics on the topic of energy-transfer dynamics in pigment-protein complexes. The convergence of these two communities has unveiled a mismatch between the background and terminology of the respective fields. To make connections, we provide a pedagogical guide to understanding the basics of two-dimensional electronic spectra aimed at researchers with a background in quantum optics.
\end{abstract}
\maketitle

\section{Introduction}
Scientists across several disciplines have recently become interested in the possibility that quantum-mechanical phenomena may play a role in the energy-transfer processes of photosynthetic organisms and synthetic light-absorbing materials. This interest was generated primarily by the observation of oscillations in cross peaks present in two-dimensional electronic spectra (2D ES) \cite{Engel:2007aa,Schlau-Cohen2009,Collini:2010aa,Panitchayangkoon:2010aa,Turner:2012ab,Schlau-Cohen2012}. \\

Oscillations during the waiting time in a 2D ES experiment represent the phase evolution of coherent superposition states generated during the experiment; more specifically, these oscillations are directly related to the phase evolution of off-diagonal elements in the density matrix, known as \emph{coherences}. Coherences can be vibrational, electronic, vibronic, etc. in character. In the simple case of a coupled heterodimer, coherent phase evolution of eigenstate superpositions is indicative of what can be thought of as \emph{coherent energy transfer} between the individual systems; a basis transformation from the energy eigenbasis of the coupled system into the subsystem basis reveals oscillating populations between the subsystems. These populations oscillate as long as the peaks in a 2D spectrum do. In particular, electronic coherences have been the focus of recent attention with respect to photosynthetic systems. Such systems are far more complicated than a coupled-heterodimer, however, it is tantalizing to ask whether oscillations observed in 2D ES are also indicative of \emph{coherent energy transfer} in photosynthetic systems \cite{Leegwater1996,Olaya-Castro2008,Nazir2009,Scholes2010}. \\

In contrast to the above discussion, conventional energy transfer between molecules is considered to be an \emph{incoherent} process. There are many senses in which the term ``incoherent'' applies in this context and, therefore, it may not be immediately obvious how to interpret the notion of \emph{coherent} energy transfer.  Before we motivate our discussion of 2D ES, we elaborate on the context in which the terminology is used. \\

We begin by considering a collection of atoms that are isolated from the environment and separated by distances smaller than the wavelength of visible light. In quantum electrodynamics, atoms couple to the vacuum modes of the universe, and an initial excitation on one or more of these atoms will result in energy transfer between them. The energy transfer is mediated via virtual photons, where the term ``virtual'' arises because such photons cannot be observed. Such a mechanism is often considered to be radiationless, however, it was shown by Andrews \emph{et al.} \cite{Andrews1989,Andrews2004} that it is simply the short-range limit of a unified theory of resonant energy transfer which includes emission of a photon by one atom followed by absorption by another. The rate of energy transfer is highest when the atoms are on resonance and drops off very quickly off-resonance \footnote{Even isolated atoms have a finite, albeit extremely narrow, linewidth due to the finite temporal width of emitted photons.}. This process is equivalent to F\"orster's phenomenological treatment of the radiationless mechanism for energy transfer \cite{Forster:1948aa} which is now known as F\"orster resonance energy transfer (FRET).\\

Interactions with an environment can lead to fluctuations in energy levels. This not only reduces the rate of energy transfer between resonant atoms, but also facilitates energy transfer between off-resonant atoms by sometimes bringing them into resonance.\\

For molecular systems, a number of additional physical processes come into play. After excitation, a molecule can lose energy to its surroundings, transitioning from a higher-energy vibrational state to a lower-energy one while still remaining in the electronic excited state. After excitation, excited energy levels can also shift due to reorganization of atomic nuclei. These processes result in a shift in the emission energy with respect to the excitation energy, collectively known as the \emph{Stokes shift}. The combined effect of these processes is that FRET between molecules tends to be unidirectional: energy can transfer from high energy molecules (known as \emph{donors}) to low energy molecules (known as \emph{acceptors}) due to high overlap between fluorescence spectra of donors and absorption spectra of acceptors; however, energy tends not to transfer from acceptors to donors due to the low overlap between fluorescence spectra of acceptors and absorption spectra of donors.\\

Despite all these nuances, the energy-transfer mechanism is always the same: energy transfer mediated via photons. Whether this is thought of as a coherent or incoherent process is largely a consequence of how terminology evolved within different disciplines. We now turn our attention to discussing such possible interpretations.\\

The most obvious way in which this type of energy transfer could be considered incoherent is due to the lack of coherence between the radiation field that excites the donor and the radiation field that is emitted by the acceptor. This can be seen even with a semiclassical treatment. If we consider the field quantum-mechanically, we can identify other possible definitions. To do so, it will be instructive to review the distinction between \emph{state coherence} and \emph{process coherence} in quantum mechanics. This was done in the introductory section of a recent paper by Kassal \emph{et al.} \cite{Kassal2013}, also in the context of energy transfer in photosynthesis.\\

A quantum state \footnote{In many fields, the term ``state'' is used synonymously with the term ``eigenstate''. Eigenstates are a property of a system determined by the system's Hamiltonian and do not change as long as the Hamiltonian remains unchanged. A quantum system, however, does not need to be in an eigenstate and can take on an arbitrary state which may consist of superpositions or mixtures of eigenstates. We use the term ``state'' in this more general context. } is described by a density matrix $\hat\rho$. Off-diagonal elements of $\hat\rho$ contain phase information and are therefore usually called ``coherences''. State coherences are basis-dependent: a state diagonal in one orthonormal basis will not be diagonal in any other. A basis-independent concept of state coherence can be defined in terms of the state purity $P=\mathrm{Tr}[\hat\rho^2]$, where a completely mixed state is proportional to the identity and contains no coherences. The state is then fully coherent when $P=1$ and fully incoherent if $P = 1/N$ where $N$ is the dimension of the system. A state can also be considered incoherent if it leads to a zero expectation value of some interesting dynamical quantity, which is directly related to whether the state is diagonal in the relevant basis. For example, the expectation value of the electric field operator with respect to a coherent state $\ket{\alpha}{}=\exp(-|\alpha|^2/2)\sum_{n}\alpha^{n}/\sqrt{n!}\ket{n}{}$ is equal to $\alpha$, while the expectation value of the electric field operator with respect to a thermal state $\hat{\rho}=\sum_{n}{P}_{n}\ket{n}{}\bra{n}{}$ is equal to zero.  During the energy-transfer process discussed above, the expectation value of the electric field is always zero, and therefore this process could be thought of as incoherent by this definition as well.\\

However, in both cases above, our conclusions of incoherence were based on the coherence properties of the \emph{state} of the radiation field and not on the energy-transfer \emph{process} itself. In quantum mechanics, the coherence of a process depends on the degree to which the evolution of an open quantum system is dominated by the unitary part or the dissipative part. In a molecular system, a process is more coherent the more strongly the molecules are coupled to each other, relative to the strength of their coupling to the environment. Using this definition of process coherence, FRET is considered incoherent because the donor and the acceptor are more strongly coupled to the environment than they are to each other, meaning that the slow transport between them can be described by rate equations. However, if two molecules are strongly coupled, unitary evolution dominates and coherent population oscillations are possible. It is in this context that the term \emph{coherent energy transfer} is intended.\\

Cross-peak oscillations in 2D spectra of light-harvesting complexes persist on time scales that indicate partial process coherence and therefore partially-coherent energy transfer. Process coherence is independent of the coherence properties of the light used to excite the system; coherent light is simply a convenient method with which to probe the coherence of the process. Incoherent light has now also been used to perform 2D ES experiments \cite{Turner:2012ad,Turner:2013aa}.\\

These measurements have motivated theoretical work on energy-transfer dynamics in the intermediate coupling regime \cite{May:2004aa,Ishizaki2009,Ishizaki2009a,Fassioli:2009aa,Roden:2009aa,Rebentrost:2009aa,Andrews:2011aa,Kolli:2011aa,Ishizaki:2011aa,Huo:2011aa,Pachon:2011aa,Miller:2012aa,Banchi2013} where the intermolecular coupling is comparable to the molecule-environment coupling, as well as quantum-information aspects such as entanglement \cite{Caruso:2009aa,Caruso:2010aa,Sarovar:2010aa,Fassioli:2010aa,CaycedoSoler:2012aa}. The convergence of nonlinear optical spectroscopy and quantum optics on the topic of energy transfer has brought to light several disparities between the conventions of the two fields. Here we provide a systematic treatment of nonlinear optical spectroscopy measurements in terminology which should be familiar to researchers in quantum optics. \\

We begin by providing a historical account of 2D ES in Section \ref{sec:history}, then review how spectroscopists understand 2D ES from a classical-physics perspective in Section \ref{sec:interpretation}, setting the scene for the semiclassical treatment of four-wave mixing used in 2D ES in Section \ref{sec:theory}. We use a heterodimer system as a model in Section \ref{sec:example} to explore basic themes of 2D ES; we include a simple phenomenological treatment of nonunitary processes such as dephasing and population relaxation, as well as data from 2D ES measurements of a GaAs quantum well sample to further illustrate these themes.  We finish in Section \ref{sec:diagrams} by describing the diagrammatic method known as double-sided Feynman diagrams  and relating it to the derivation in previous sections.

\section{History of 2D ES}
\label{sec:history}
Coherent multidimensional optical spectroscopy---of which 2D ES is but one embodiment---is a type of nonlinear optical spectroscopy measurement.  For broad treatments of nonlinear optics and nonlinear optical spectroscopy methods, we refer readers to texts including those by Boyd \cite{Boyd:2003aa} and Mukamel \cite{Mukamel:1995aa}. We refer readers to the following texts \cite{Cho:2009aa,Hamm:2011aa} and review articles \cite{Jonas:2003aa,Cho:2008aa,Abramavicius:2009aa,Lewis:2012aa,Cundiff2013} for treatments more specific to coherent multidimensional optical spectroscopy. \\

2D ES was developed in the late 1990s \cite{Hybl:1998aa,Hybl:2001aa,Jonas:2003aa}, concurrently with 2D IR \cite{Hamm:1998aa,Zanni:2001aa,Demirdoven:2002aa}. Many aspects of the technique are related to nuclear magnetic resonance (NMR) methods developed decades earlier \cite{Ernst:1987aa,Jonas:2003ab}. 2D ES has now been used to study a variety of systems including atoms \cite{Tian:2003aa,Dai:2010aa,Li:2013aa}, molecules \cite{Brixner:2004aa,Egorova:2008aa,Christensson:2009aa,Prokhorenko:2009aa,Nemeth:2010aa,Nemeth:2010ab,Fidler:2010aa,Christensson:2011aa,Turner:2011ab,Harel:2011aa}, molecular aggregates \cite{Ginsberg:2009aa,Arias:2013aa} and related nanostructures \cite{Milota:2009aa,Nemeth:2009ab,Wen:2013aa}, biological pigment-protein complexes \cite{Brixner:2005aa,Engel:2007aa,Anna2009,Womick:2009aa,Collini:2010aa,Ginsberg:2011aa,Harel:2011ab,Lewis:2012aa,Turner:2012ab,Anna2013,Ostroumov2013}, and semiconductor nanostructures \cite{Borca:2005aa,Li:2006gf,Zhang:2007zr,Stone:2009rt,Bristow:2009ab,Cundiff:2009aa,Turner:2010aa,Wong:2011ab,Davis:2011aa,Turner:2012aa,Turner:2012ac}. Methods for measuring multidimensional spectra using entangled photons have been proposed \cite{Roslyak:2009aa}. The method has found widespread adoption because it gives information beyond that which can be obtained from more conventional femtosecond transient-absorption measurements \cite{Brito-Cruz:1986aa,Dantus:1987aa,Ruhman:1987aa,Chesnoy:1988aa,Walmsley:1989aa,Fragnito:1989aa,Pollard:1990aa,Dantus:1990aa,Potter:1992aa,Scherer:1993aa,Vos:1993aa,Wang:1994aa,Mizutani:1997aa,Assion:1998aa,Meyer:1998aa,Shah:1999aa,Shen:1999aa,Fuji:2000aa,Chemla:2001aa,Prokhorenko:2002aa,Herek:2002aa,Sewall:2006aa,Sagar:2008ab,Polli:2010aa,Kambhampati:2011aa}. \\

The main advantages of 2D ES are that it can separate inhomogeneous and homogeneous line-broadening contributions and it can expose microscopic couplings directly through the presence of cross peaks. The observation of cross peaks---and oscillating cross peaks in particular---in 2D ES measurements of pigment-protein complexes has been extensively discussed \cite{Brixner:2005aa,Egorova:2007aa,Egorova:2008aa,Nemeth:2008aa,Mancal:2010aa,Nemeth:2010ab,Christensson:2011aa,Turner:2011ab,Hayes:2011aa,Turner:2012ab,Richards:2012aa,Tiwari:2013aa}.  2D ES is becoming more widespread, and claims arising from 2D ES measurements have led to interest by the quantum-optics community.

\section{Classical interpretation}
\label{sec:interpretation}
The way in which many spectroscopists understand four-wave mixing signals is based on classical optics concepts including interference patterns and diffraction.\\

Before discussing 2D ES, we begin with the more intuitive transient-absorption spectroscopy, a type of pump-probe spectroscopy. This technique is one type of four-wave mixing measurement in which only two femtosecond pulses are used. A pump pulse excites the sample, and after a delay, $\tau_2$, a weaker probe pulse measures the result. One can obtain spectral information about the pump-induced absorption change by frequency-resolving the probe; one can also obtain temporal information by repeating the measurement at many values of $\tau_2$.  Formally, the pump pulse contributes two field-matter interactions simultaneously (one with positive wave vector and the other with negative wave vector) such that the total signal is emitted in the $\bf{k}_{\mathrm{sig}} = \bf{k}_{\mathrm{pump}} - \bf{k}_{\mathrm{pump}} + \bf{k}_{\mathrm{probe}} =  \bf{k}_{\mathrm{probe}}$ direction. The wave vector $\mathbf{k}$ is related to wavelength of light via $|\mathbf{k}|=2\pi/\lambda$.\\

\begin{figure}
\centering
  \includegraphics[width=0.75\textwidth]{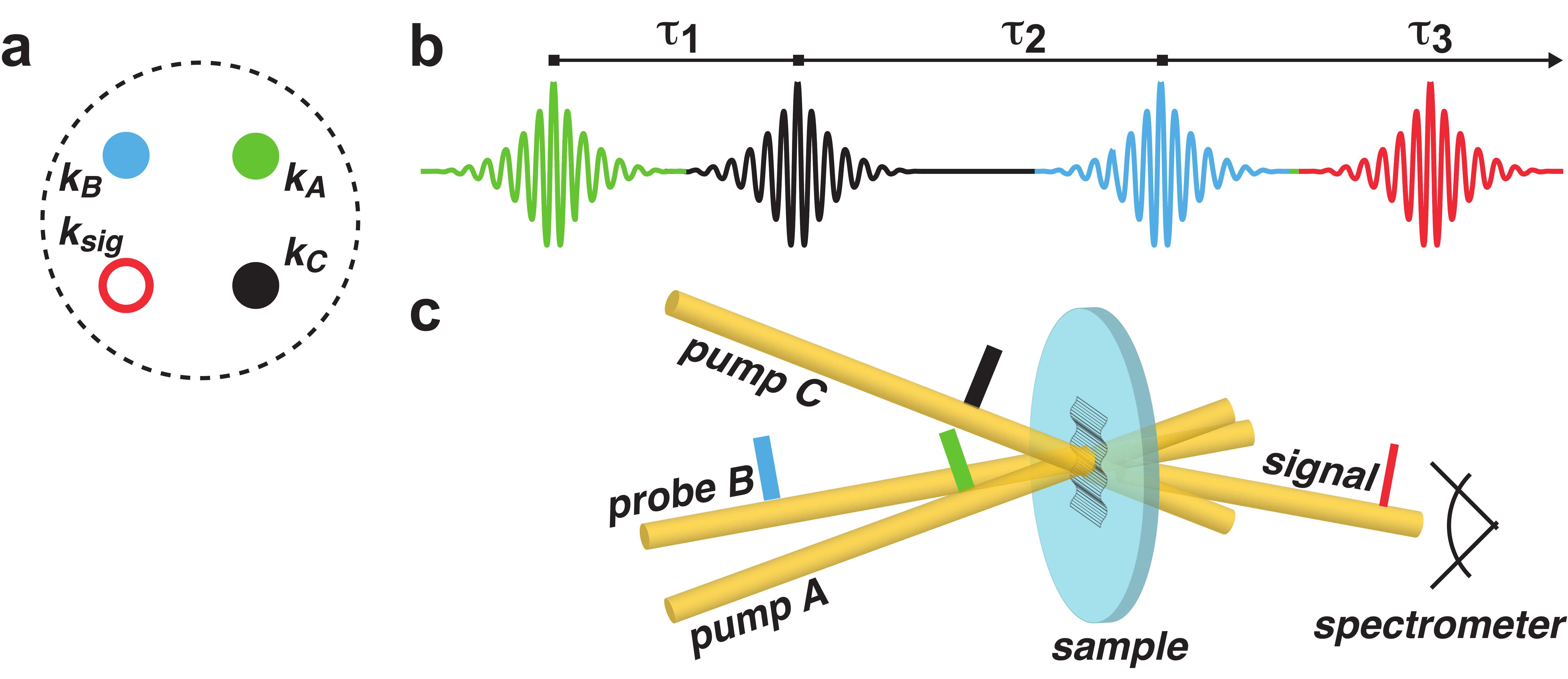} 
  \caption{Parameters of a four-wave mixing measurement using femtosecond laser pulses. (a) Confocal view of the BOX beam pattern. Because we chose the red beam to indicate the signal direction, the green beam is therefore the beam which contributes a negative wave vector, $-{\bf k}$, to the signal. The blue and black beams each contribute positive wave vector to the signal, ${\bf k}_{\mathrm{sig}} = -{\bf k}_{\mathrm{A}} + {\bf k}_{\mathrm{B}} + {\bf k}_{\mathrm{C}}$. (b) The rephasing pulse-timing sequence because the green beam interacts first. (c) The interference pattern caused by the two pump pulses induces a grating in the sample.  The signal is the portion of the probe pulse that is diffracted by the induced grating into the direction that conserves energy and momentum.}
  \label{fig:signal_generation}
\end{figure}

In many ways, the common implementation of 2D ES is an extension of transient absorption. The method not only allows access to the same temporal and emission spectral information contained in a transient-absorption spectrum, but also it allows access to the excitation spectrum. This leads to a 2D `map' correlating excitation and emission frequencies for each $\tau_2$ delay value. The main technical distinction between transient absorption and 2D ES is that the single pump pulse is replaced by a pair of pump pulses. By stepping the delay between the pair of pump pulses followed by Fourier transformation with respect to the time delay between them, one can work out the excitation spectrum.  We depict the signal-generation process in Fig. \ref{fig:signal_generation}. The input beams are arranged on the corners of a square. This is called the BOX geometry, from the term BOXCARS used in many coherent anti-Stokes Raman scattering experiments. With three intense femtosecond pulses, many optical effects can be produced. Because of the pulse energies used and the detection of signal in a specific, phase-matched direction, $\bf{k}_{\mathrm{sig}} = -\bf{k}_{\mathrm{A}} + \bf{k}_{\mathrm{B}} + \bf{k}_{\mathrm{C}}$, the femtosecond pulse in each beam contributes one field-matter interaction to the signal-generation process. The fourth field-matter interaction is the signal emission. Both transient absorption and 2D ES use a four-wave mixing process, which involves the third-order nonlinear optical susceptibility $\chi^{(3)}$. (Higher-order nonlinearities have been used in 2D ES measurements as well \cite{Fulmer:2005aa,Turner:2010aa,Turner:2011aa,Turner:2011ad,Fidler:2010aa}. Here we focus on signals from third-order nonlinearities.) We label each beam in Fig. \ref{fig:signal_generation} with letters to avoid confusing beam directions (which do not change in the experimental implementation because one typically measures only one signal direction) with relative time orderings (which can vary depending on the chosen pulse sequence).  In Fig. \ref{fig:signal_generation}, the two pulses that form the pump sequence are coloured green (contributing $-{\bf k}$) and black (contributing $+{\bf k}$). In the limit that the two pump pulses converge (${\bf k}_{\mathrm{A}} = {\bf k}_{\mathrm{B}}$, and consequently ${\bf k}_{\mathrm{sig}} = {\bf k}_{\mathrm{C}}$), we recover the transient-absorption measurement. It is possible---and sometimes advantageous---to perform 2D ES in the transient-absorption geometry \cite{Grumstrup:2007aa,Deflores:2007aa,Shim:2009aa,Tekavec:2009aa}.\\

Typically in 2D ES the two pump beams cross at a small angle \cite{Yetzbacher:2007aa}. When the relative optical phase between the two pulses is stable, they form a stable interference pattern in the electromagnetic field as depicted in Fig. \ref{fig:signal_generation}c by the wavy black line on the sample surface. The spacing of this interference fringe is typically several micrometers, and photographs of the grating are available \cite{Maznev:1998aa,Bristow:2008aa}. The periodicity of the fringe is given by $\Lambda = \lambda / (2 \sin(\theta/2))$ where  $\lambda$ is the excitation wavelength and $\theta$ is the crossing angle between the two pump beams \cite{Slayton:2001aa}. This interference pattern is imprinted, or `written', into the sample when the electromagnetic field interacts with the bound electrons of the material to induce a spatially periodic polarization. The grating pattern is then probed by the third pulse, coloured blue in Fig. \ref{fig:signal_generation}. The signal field is the portion of the probe pulse that is diffracted by the induced grating into the phase-matched direction. \\

In this paper, we focus on modelling diffraction due to the electronic grating and how this diffraction changes as a function of the pulse timing adjustments and sample parameters. Often the electronic excitation leads to nuclear motion, resulting from one or more physical processes. In impulse-stimulated thermal scattering, the bright areas of the sample absorb light which gets converted into heat during radiationless relaxation. The temperature of the medium therefore becomes modulated across the grating resulting in a long-lived refractive index grating.  In addition, a difference-frequency mixing process can excite coherent vibrational oscillations in what is known as impulsive-stimulated Raman scattering.\\

For simple systems, as the time delay between the pumps and the probe (again denoted $\tau_{2}$ to be consistent with the transient-absorption measurement) increases, the amount of light diffracted into the signal direction decreases exponentially as the kinetics of the sample decay the system back to the ground state. Moreover, for a given $\tau_{2}$ value, as the time delay between the two pump pulses, $\tau_{1}$, is varied, the phase and amplitude of the imprinted grating pattern will change. In the limit where the two pump pulses are further apart than the duration of the optical polarization induced in the material by each pulse (which can be longer than the pulse duration), no interference occurs and therefore no grating pattern is induced. For small $\tau_{1}$ values, the phase of the interference pattern changes, and this phase shift can be measured if one can resolve the phase of the emitted signal field. This is a key feature of 2D ES measurements. Often, the emitted signal is measured using a spectrometer and CCD to spectrally resolve the field, $E^{(3)}_{\mathrm{sig}}(\omega_3)$, and a fully characterized reference field (often called a local-oscillator) is used to assist in determining the phase of the emitted signal \cite{Lepetit:1995aa,Dorrer:1999ab}. \\

The description above can be used to understand the full range of third-order nonlinear spectroscopy experiments including transient-absorption, transient-grating, self-diffraction, three-pulse photo-echo peak-shift, impulsive-stimulated scattering, and 2D optical spectroscopy.  The different experiments are just different pulse-timing sequences or projections through the three dimensions of the signal, $E^{(3)}_{\mathrm{sig}}(\tau_{1},\tau_{2},\tau_{3})$. In most cases, because the dynamics during time period $\tau_{2}$ are either near zero frequency (for population decays) or low frequencies, that dimension is usually displayed in the time domain. The coherent oscillations during time periods $\tau_{1}$ and $\tau_{3}$ are much faster in comparison, at optical frequencies, so those are usually displayed in the frequency domain. In other words, the signal is usually displayed as some projection of $E^{(3)}_{\mathrm{sig}}(\omega_1,\tau_{2},\omega_3)$. \\

When a sequence of intense femtosecond pulses interacts with the sample, a great variety of events can transpire. We are interested in the small subset of events that leads to coherent signal generated in the phase-matched direction by the induced electronic grating. We then use this signal---under certain pulse-timing sequences---to produce the 2D electronic spectrum.

\section{Semiclassical theory}
\label{sec:theory}
A semiclassical treatment is often used to describe the observed physical phenomenon discussed above. We follow the method developed primarily by Mukamel and coworkers \cite{Mukamel:1995aa}. \\

In the semiclassical method, the light is treated classically while the material---characterized by the system Hamiltonian $\ham_{\textsc{s}}$---is treated quantum mechanically. The treatment of the interaction between the laser pulses and the system is split into two parts. First, the nonlinear polarization $P(\mathbf{r},t)$ is calculated by evolving the  state of the material system according to the light-matter interaction Hamiltonian $\ham_{\textsc{lm}}(\mathbf{r},t)=\hat{\mu}\cdot{E}_{\textsc{t}}(\mathbf{r},t)$ in the dipole approximation, where $\hat{\mu}$ is the transition dipole moment operator and ${E}_{\textsc{t}}(\mathbf{r},t)$ is the total electric field describing all incoming pulses.  Second, the nonlinear polarization $P(\mathbf{r},t)$ is used as a source term in the electromagnetic wave equation to calculate the generated signal field $E_{\mathrm{sig}}(\mathbf{r},t)$.  For simplicity we disregard the orientations of the field polarization vectors and transition dipoles, retaining only their relative amplitudes. Below we describe the process, starting from the third-order signal field, $E^{(3)}_{\mathrm{sig}}$ and working through the third-order polarization, $P^{(3)}_{\mathrm{sig}}$, to the time-dependent system density matrix, $\hat{\rho}^{(3)}(t)$. \\

\subsection{The electromagnetic wave equation}
In a 2D ES measurement derived from a third-order optical nonlinearity, three electric fields delivered by femtosecond laser pulses induce a nonlinear polarization $P(t)$ in the sample which emits a new coherent beam in the phase-matched direction. To predict a 2D spectrum, we therefore must describe how laser pulses interact with the system. This interaction is given by the wave equation in nonlinear optical media \cite{Boyd:2003aa}:
\begin{align}\label{eq:wave}
\nabla^2 E(t)-\frac{n^2}{c^2}\frac{\partial^2 E(t)}{\partial t^2}={}&\frac{1}{\epsilon_0c^2}\frac{\partial^2 P(t)}{\partial t^2}\,,
\end{align}
where $n$ is the linear refractive index, $c$ is the speed of light in vacuum and $\epsilon_0$ is the permittivity of free space.\\

If one assumes the semi-impulsive limit---where the laser pulses are short compared with any time scale of the system but long compared to the oscillation period of the light field---as well as a sufficiently weak generated signal to avoid back action, the solution to Eq. (\ref{eq:wave}) will be $\pi/2$ out-of-phase with the polarization of the sample, meaning
\begin{align}\label{eq:Esigproto}
E^{(+)}_{\mathrm{sig}}(t)\propto i P^{(+)}(t)\,,
\end{align}
where $E^{(+)}_{\mathrm{sig}}$ and  $P^{(+)}(t)$ are the positive-frequency parts of the field and polarization respectively. To determine the signal field, we thus need to compute the nonlinear polarization. \\

\subsection{Calculating the nonlinear polarization}
\label{sec:resp}
The nonlinear polarization is a macroscopic collective dipole moment per unit volume and can be written  as the expectation value of the transition dipole moment operator $\hat{\mu}$,
\begin{align}
P(t)=\langle \hat{\mu}\rangle=\mathrm{Tr}[\hat{\mu}\hat{\rho}(t)]\,,
\end{align}
where $\hat{\rho}(t)$ is the state of the material system at time $t$ and evolves according to system and light-matter Hamiltonians $\ham_{\textsc{s}}$ and $\ham_{\textsc{lm}}(t)$ respectively. It is convenient to work in the interaction picture, within which the polarization becomes
\begin{align}\label{eq:pol3}
P(t)=\mathrm{Tr}[\hat{\mu}_{\textsc{i}}(t)\hat{\rho}_{\textsc{i}}(t)]\,,
\end{align}
where 
\begin{align}\label{eq:mut}
\hat{\mu}_{\textsc{i}}(t)={}&\ee{+\tfrac{i}{\hbar}\ham_{\textsc{s}}t}\hat{\mu}\ee{-\tfrac{i}{\hbar}\ham_{\textsc{s}}t}\,,
\end{align}
is the transition dipole moment operator in the interaction picture. The light-matter interaction Hamiltonian becomes
\begin{align}
\ham_{\textsc{i}}(t)={}&\ee{+\tfrac{i}{\hbar}\ham_{\textsc{s}}t}\ham_{\textsc{lm}}(t)\ee{-\tfrac{i}{\hbar}\ham_{\textsc{s}}t}\\\label{eq:Hamthing}
={}&\hat{\mu}_{\textsc{i}}(t)\cdot E_{\textsc{t}}({t})\,,
\end{align}
and the density operator $\hat{\rho}_{\textsc{i}}(t)$ now evolves under $\ham_{\textsc{i}}(t)$ according to
\begin{align}\label{eq:rhot}
\hat{\rho}_{\textsc{i}}(t)=\mathcal{T}\ee{-\tfrac{i}{\hbar}\int_{{t}_{0}}^{t}\ham_{\textsc{i}}({t}')d{t}'}\hat{\rho}_{0}\ee{+\tfrac{i}{\hbar}\int_{{t}_{0}}^{t}\ham_{\textsc{i}}({t}')d{t}'}\,,
\end{align}
where $\hat{\rho}_{0}$ is the initial state of the molecular system in the interaction picture and $\mathcal{T}$ is the time-ordering operator. \\

We assume that $\ham_{\textsc{lm}}$ is sufficiently weak to justify a perturbative approach. Since we are interested in a third-order nonlinear effect, the relevant term of the formal solution to Eq. (\ref{eq:rhot}) is 
\begin{align}\label{eq:rho3}
\begin{split}
\hat{\rho}^{(3)}({t})={}&\left(-\frac{i}{\hbar}\right)^3\int_{{t}_{0}}^{t}d{t}'_{3}\int_{{t}_{0}}^{{t}'_{3}}d{t}'_{2}\int_{{t}_{0}}^{{t}'_{2}}d{t}'_{1}[\ham_{\textsc{i}}({t}'_{3}),[\ham_{\textsc{i}}({t}'_{2}),[\ham_{\textsc{i}}({t}'_{1}),\hat{\rho}_{0}]]]\,.
\end{split}
\end{align}

Inserting Eq. (\ref{eq:rho3}) into Eq. (\ref{eq:pol3}) gives the third-order nonlinear polarization 

\begin{align}\label{eq:pol4}
\begin{split}
P^{(3)}(t)={}&\left(-\frac{i}{\hbar}\right)^3\int_{{t}_{0}}^{t}d{t}'_{3}\int_{{t}_{0}}^{{t}'_{3}}d{t}'_{2}\int_{{t}_{0}}^{{t}'_{2}}d{t}'_{1} E_{\textsc{t}}({t}'_{3})E_{\textsc{t}}({t}'_{2})E_{\textsc{t}}({t}'_{1})\mathrm{Tr}[\hat{\mu}_{\textsc{i}}(t)[\hat{\mu}_{\textsc{i}}({t}'_{3}),[\hat{\mu}_{\textsc{i}}({t}'_{2}),[\hat{\mu}_{\textsc{i}}({t}'_{1}),\hat{\rho}_{0}]]]]\,.
\end{split}
\end{align}

The next step is to consider the details of the incoming field, which will consist of three laser pulses
\begin{align}\label{eq:Efield4}
E_{\textsc{t}}(t)=E_1(t)+E_2(t)+E_3(t)\,,
\end{align}
with arrival times for pulse $E_{j}(t)$ centred around ${T}_{j}$. Since each pulse can be decomposed into positive and negative frequency terms $E_j(t)=E^+_j(t)+E^-_j(t)$, Eq. (\ref{eq:Efield4}) will contain six terms. In turn, the product $E_{\textsc{t}}({t}'_{3})E_{\textsc{t}}({t}'_{2})E_{\textsc{t}}({t}'_{1})$ within Eq. (\ref{eq:pol4}) will consist of $6^3=216$ terms in total!  Fortunately, we can make an approximation to reduce the number of terms. In the semi-impulsive limit, 
\begin{align}\label{eq:imp}
E^+_{j}({t})\approx{}&E_{j}\delta({t}-{T}_{j})\ee{+i(\bar{\omega}_{j} {t}-\mathbf{k}_{j}\cdot\mathbf{r})},
\end{align}
and we can impose strict time ordering ${T}_{1}<{T}_{2}<{T}_{3}$. Since this limit is only \emph{semi}-impulsive, the expressions for the fields still contain carrier frequencies ($\bar{\omega}_{j}$) to account for pulses with finitely broad spectra. The expression simplifies significantly to 
\begin{align}\label{eq:aneq}
\begin{split}
P^{(3)}(t)={}&\left(-\frac{i}{\hbar}\right)^3\int_{{t}_{0}}^{t}d{t}'_{3}\int_{{t}_{0}}^{{t}'_{3}}d{t}'_{2}\int_{{t}_{0}}^{{t}'_{2}}d{t}'_{1} E_{3}({t}'_{3})E_{2}({t}'_{2})E_{1}({t}'_{1})\mathrm{Tr}[\hat{\mu}_{\textsc{i}}(t)[\hat{\mu}_{\textsc{i}}({t}'_{3}),[\hat{\mu}_{\textsc{i}}({t}'_{2}),[\hat{\mu}_{\textsc{i}}({t}'_{1}),\hat{\rho}_{0}]]]]\,,
\end{split}
\end{align}
with only $2^3=8$ terms from the field contributions. The difference between Eqs.  (\ref{eq:pol4}) and (\ref{eq:aneq}) is subtle, and we therefore direct the reader toward the subscripts on the electric fields. \\

In the remaining steps, we perform some mathematical acrobatics to write the polarization in terms of time delays between the pulses. We first use the cyclic properties of the trace to write
\begin{align}
\begin{split}
P^{(3)}(t)={}&\left(-\frac{i}{\hbar}\right)^3\int_{{t}_{0}}^{t}d{t}'_{3}\int_{{t}_{0}}^{{t}'_{3}}d{t}'_{2}\int_{{t}_{0}}^{{t}'_{2}}d{t}'_{1} E_{3}({t}'_{3})E_{2}({t}'_{2})E_{1}({t}'_{1})\mathrm{Tr}[[[[\hat{\mu}_{\textsc{i}}(t),\hat{\mu}_{\textsc{i}}({t}'_{3})],\hat{\mu}_{\textsc{i}}({t}'_{2})],\hat{\mu}_{\textsc{i}}({t}'_{1})]\hat{\rho}_{0}]\,.
\end{split}
\end{align}

We then send ${t}_{0}\rightarrow -\infty$ and perform a change of variable to de-correlate the time integrals: ${t}'_{1}\rightarrow {t}-{t}_{1}-{t}_{2}-{t}_{3}$, ${t}'_{2}\rightarrow {t}-{t}_{2}-{t}_{3}$ and ${t}'_{3}\rightarrow {t}-{t}_{3}$, which gives
\begin{align}
P^{(3)}(t)={}&\left(\frac{i}{\hbar}\right)^3\int_{0}^{\infty}d{t}_{3}\int_{0}^{\infty}d{{t}}_{2}\int_{0}^{\infty}d{{t}}_{1} E_{3}({t}-{t}_{3})E_{2}({t}-{t}_{2}-{t}_{3})E_{1}({t}-{t}_{1}-{t}_{2}-{t}_{3})\\\nonumber
&\times\mathrm{Tr}[[[[\hat{\mu}_{\textsc{i}}(t),\hat{\mu}_{\textsc{i}}({t}-{t}_{3})],\hat{\mu}_{\textsc{i}}({t}-{t}_{2}-{t}_{3})],\hat{\mu}_{\textsc{i}}({t}-{t}_{1}-{t}_{2}-{t}_{3})]\hat{\rho}_{0}]\,.
\end{align}

We use the definition of the time-dependent dipole moment operator in Eq. (\ref{eq:mut})---and the fact that $\hat{\rho}_{0}$ does not evolve under $\ham_{\mathrm{s}}$---to rewrite the argument of the trace. This amounts to adding ${t}_{1}+{t}_{2}+{t}_{3}-{t}$ to all the time arguments in the trace, yielding
\begin{align}
P^{(3)}(t)={}&\left(\frac{i}{\hbar}\right)^3\int_{0}^{\infty}d{t}_{3}\int_{0}^{\infty}d{{t}}_{2}\int_{0}^{\infty}d{{t}}_{1} E_{3}({t}-{t}_{3})E_{2}({t}-{t}_{2}-{t}_{3})E_{1}({t}-{t}_{1}-{t}_{2}-{t}_{3})\\\nonumber
&\times\mathrm{Tr}[[[[\hat{\mu}_{\textsc{i}}({t}_{1}+{t}_{2}+{t}_{3}),\hat{\mu}_{\textsc{i}}({t}_{1}+{t}_{2})],\hat{\mu}_{\textsc{i}}({t}_{1})],\hat{\mu}_{\textsc{i}}(0)]\hat{\rho}_{0}]\,.
\end{align}

We now evaluate the time integrals---which are straightforward given the fields in Eq. (\ref{eq:imp})---and write the polarization explicitly in terms of the time delays between the three pulses:
\begin{align}\label{eq:some}
P^{(3)}(\tau_{1},\tau_{2},\tau_{3})={}&\left(\frac{i}{\hbar}\right)^3\theta(\tau_{1}) \theta(\tau_{2}) \theta(\tau_{3})\mathcal{E}_1\mathcal{E}_2\mathcal{E}_3\mathrm{Tr}[[[[\hat{\mu}_{4},\hat{\mu}_{3}],\hat{\mu}_{2}],\hat{\mu}_{1}]\hat{\rho}_{0}]\,,
\end{align}
where $ \mathcal{E}_{j}=\mathcal{E}^+_{j}+\mathcal{E}^-_{j}$ and
\begin{align}\label{eq:tildeE}
\mathcal{E}^+_{j}={}&E_{j}e^{+i (\mathbf{k}_{j}\cdot\mathbf{r}-\bar{\omega}_{j}  {T}_{j})}\,,
\end{align}
is the positive \emph{momentum} component of $\mathcal{E}_{j}$, and where $\theta [ \ldots ]$ represents the Heaviside step function. The transition dipoles are defined to be 
\begin{align} 
\hat{\mu}_{1}={}&\hat{\mu}_{\textsc{i}}(0)\\
\hat{\mu}_{2}={}&\hat{\mu}_{\textsc{i}}(\tau_{1})\\
\hat{\mu}_{3}={}&\hat{\mu}_{\textsc{i}}(\tau_{1}+\tau_{2})\\
\hat{\mu}_{4}={}&\hat{\mu}_{\textsc{i}}(\tau_{1}+\tau_{2}+\tau_{3})\,,
\end{align}
where $\tau_{3}={t}-{T}_{3}$, $\tau_{2}={T}_{3}-{T}_{2}$ and $\tau_{1}={T}_{2}-{T}_{1}$ are time delays between pulses, as shown in Fig. \ref{fig:signal_generation}. \\

Eq. (\ref{eq:some}) is sufficient for calculating the total generated four-wave mixing signal according to the expression for the electric field given in Eq. \ref{eq:Esigproto}. In the next sections we make the rotating wave approximation and show how one can use phase matching---conservation of energy and momentum---to select different parts of the signal.

\subsection{The rotating wave approximation and phase matching}
One of the advantages of 2D ES over conventional transient-absorption spectroscopy is the ability to select different parts of the total signal by controlling the properties of the incoming pulses, most often the relative time ordering. This is afforded by the noncollinear beam geometry which makes use of phase matching. \\

Given that the carrier frequencies of the pulses are comparable to the transition energies in the material system, we can invoke the rotating wave approximation (RWA), which eliminates the quickly oscillating frequency terms under the time integral. \\

We begin by decomposing the transition dipole moment operator into positive and negative frequency parts $\hat{\mu}_{\textsc{i}}({t}_{j})=\hat{\mu}^+_{\textsc{i}}({t}_{j})+\hat{\mu}^-_{\textsc{i}}({t}_{j})$. Only products of the form $E^{\mp}_{j}({t}_{j})\hat{\mu}^{\pm}_{\textsc{i}}({t}_{j})$ survive the time integrals performed in Eq. (\ref{eq:aneq}). This amounts to only keeping products $\mathcal{E}_{j}^{\pm}\hat{\mu}_{j}^{\pm}$ in Eq. (\ref{eq:some}). Making the RWA, we rewrite the nonlinear polarization by expanding the nested commutator, to give
\begin{align}\label{eq:resp}
P^{(3)}(\tau_{1},\tau_{2},\tau_{3})={}&\left(\frac{i}{\hbar}\right)^3\theta(\tau_{1}) \theta(\tau_{2}) \theta(\tau_{3})\sum_{{p}_{1}=\pm}\sum_{{p}_{2}=\pm}\sum_{{p}_{3}=\pm}\sum_{r=1}^{4}\left(F_{r}^{{p}_{1}{p}_{2}{p}_{3}}(\tau_{1},\tau_{2},\tau_{3})-\cc)\right)\,,
\end{align}
where summations over ${p}_{1}$, ${p}_{2}$ and ${p}_{3}$ are summations over labels  $+$ and $-$, e.g.  $F_{1}^{+-+}(\tau_{1},\tau_{2},\tau_{3})$ is one term in Eq.  (\ref{eq:resp}). The terms in the summation over $r$ are
\begin{subequations}\label{eq:respJ}
\begin{align}\label{eq:resp7}
F_{1}^{{p}_{1}{p}_{2}{p}_{3}}(\tau_{1},\tau_{2},\tau_{3})={}&\mathcal{E}^{{p}_{1}}_{1}\mathcal{E}^{{p}_{2}}_{2}\mathcal{E}^{{p}_{3}}_{3}\mathrm{Tr}[\hat{\mu}_{4}\hat{\mu}_{3}^{{p}_{3}}\hat{\mu}_{2}^{{p}_{2}}\hat{\mu}_{1}^{{p}_{1}}\hat{\rho}_{0}]\\
F_{2}^{{p}_{1}{p}_{2}{p}_{3}}(\tau_{1},\tau_{2},\tau_{3})={}&-\mathcal{E}^{{p}_{1}}_{1}\mathcal{E}^{{p}_{2}}_{2}\mathcal{E}^{{p}_{3}}_{3}\mathrm{Tr}[\hat{\mu}_{4}\hat{\mu}_{3}^{{p}_{3}}\hat{\mu}_{1}^{{p}_{1}}\hat{\rho}_{0}\hat{\mu}_{2}^{{p}_{2}}]\\
F_{3}^{{p}_{1}{p}_{2}{p}_{3}}(\tau_{1},\tau_{2},\tau_{3})={}&-\mathcal{E}^{{p}_{1}}_{1}\mathcal{E}^{{p}_{2}}_{2}\mathcal{E}^{{p}_{3}}_{3}\mathrm{Tr}[\hat{\mu}_{4}\hat{\mu}_{2}^{{p}_{2}}\hat{\mu}_{1}^{{p}_{1}}\hat{\rho}_{0}\hat{\mu}_{3}^{{p}_{3}}]\\
F_{4}^{{p}_{1}{p}_{2}{p}_{3}}(\tau_{1},\tau_{2},\tau_{3})={}&\mathcal{E}^{{p}_{1}}_{1}\mathcal{E}^{{p}_{2}}_{2}\mathcal{E}^{{p}_{3}}_{3}\mathrm{Tr}[\hat{\mu}_{4}\hat{\mu}_{1}^{{p}_{1}}\hat{\rho}_{0}\hat{\mu}_{2}^{{p}_{2}}\hat{\mu}_{3}^{{p}_{3}}]\,.
\end{align}
\end{subequations}

The electric field will be $\pi/2$ out-of-phase with the non-linear polarization of the sample, 
\begin{align}\label{eq:resp}
\varepsilon^{(3)}(\tau_{1},\tau_{2},\tau_{3})\propto{}&\theta(\tau_{1}) \theta(\tau_{2}) \theta(\tau_{3})\sum_{{p}_{1}=\pm}\sum_{{p}_{2}=\pm}\sum_{{p}_{3}=\pm}\sum_{r=1}^{4}\left(F_{r}^{{p}_{1}{p}_{2}{p}_{3}}(\tau_{1},\tau_{2},\tau_{3})+\cc)\right)\,.
\end{align}

The real electric field $\varepsilon({t})$ underlying an optical pulse is twice the real part of its analytic signal ${E}({t})$: $\varepsilon({t})=2\mathrm{Re}[{E}({t})]$. Because the four-wave mixing signal leading to the 2D spectrum is almost always detected using a spectrometer, only the positive frequency components---i.e. the analytic signal---of the field are required. We can determine the analytic signal by taking the single-sided inverse of the Fourier transform of the field
\begin{subequations}
\begin{align}
E({t})={}\int_{0}^{\infty}d\omega \tilde\varepsilon(\omega)\ee{-i\omega{t}}\,;\hspace{1cm}\mathrm{where}\hspace{1cm}\tilde\varepsilon(\omega)={}\int_{-\infty}^{\infty}d{t} \varepsilon({t})\ee{i\omega{t}}\,.
\end{align}
\end{subequations}

If we assume a BOX beam geometry and are only interested in the signal emitted in the direction specified in Fig. \ref{fig:signal_generation}, then the phase matching condition that must be satisfied is $\mp\textbf{k}_{\mathrm{A}}\pm\textbf{k}_{\mathrm{B}}\pm\textbf{k}_{\mathrm{C}}\mp\textbf{k}_{\mathrm{sig}}=0$, where the labels are given in Fig. \ref{fig:signal_generation}. This eliminates the terms $\mathcal{E}^+_1 \mathcal{E}^+_2 \mathcal{E}^+_3$ and $\mathcal{E}^-_1 \mathcal{E}^-_2 \mathcal{E}^-_3$---and therefore, the terms $F_{r}^{+++}(\tau_{1},\tau_{2},\tau_{3})$ and $F_{r}^{---}(\tau_{1},\tau_{2},\tau_{3})$--- from  Eq. (\ref{eq:resp}). Furthermore, if the initial state of the system is the ground state $\hat{\rho}_{0}=\ket{g}{}\bra{g}{}$, only the terms with $\hat{\mu}^+_{j}$ ($\hat{\mu}^-_{j}$) acting from the left (right) of $\hat{\rho}_{0}$ survive, i.e. ${s}_{1}=+$.\\

By experimentally controlling the order of the incoming pulses, we can measure different parts of the total signal in Eq. (\ref{eq:resp}). Recall that we imposed strict time ordering ${T}_{1}<{T}_{2}<{T}_{3}$ after Eq. (\ref{eq:imp}). If for example pulse $A$ arrives first, e.g. ${T}_{\mathrm{A}}<{T}_{\mathrm{B}}<{T}_{\mathrm{C}}$, the only terms to survive the phase-matching condition are
\begin{subequations}\label{eq:respR}
\begin{align}\label{eq:respRd}
F_{1}^{-++}(\tau_{1},\tau_{2},\tau_{3})^*={}&\mathcal{E}^{-}_{1}\mathcal{E}^{+}_{2}\mathcal{E}^{+}_{3}\mathrm{Tr}[\hat{\rho}_{0}\hat{\mu}_{1}^-\hat{\mu}_{2}^+\hat{\mu}_{3}^+\hat{\mu}_{4}]\\
F_{2}^{-++}(\tau_{1},\tau_{2},\tau_{3})^*={}&-\mathcal{E}^{-}_{1}\mathcal{E}^{+}_{2}\mathcal{E}^{+}_{3}\mathrm{Tr}[\hat{\mu}_{2}^+\hat{\rho}_{0}\hat{\mu}_{1}^-\hat{\mu}_{3}^+\hat{\mu}_{4}]\\
F_{3}^{-++}(\tau_{1},\tau_{2},\tau_{3})^*={}&-\mathcal{E}^{-}_{1}\mathcal{E}^{+}_{2}\mathcal{E}^{+}_{3}\mathrm{Tr}[\hat{\mu}_{3}^+\hat{\rho}_{0}\hat{\mu}_{1}^-\hat{\mu}_{2}^+\hat{\mu}_{4}]\\
F_{4}^{-++}(\tau_{1},\tau_{2},\tau_{3})^*={}&\mathcal{E}^{-}_{1}\mathcal{E}^{+}_{2}\mathcal{E}^{+}_{3}\mathrm{Tr}[\hat{\mu}_{3}^+\hat{\mu}_{2}^+\hat{\rho}_{0}\hat{\mu}_{1}^-\hat{\mu}_{4}]\,,
\end{align}
\end{subequations}
where here, the analytic signal is given by only the conjugate terms in Eq (\ref{eq:resp}). The above pulse ordering can lead to a photon echo and is therefore known as a \emph{rephasing} experiment. If, on the other hand, pulse $A$ arrives second, e.g. ${T}_{\mathrm{B}}<{T}_{\mathrm{A}}<{T}_{\mathrm{C}}$, the only terms to survive the phase-matching condition are
\begin{subequations}\label{eq:respnR}
\begin{align}
F_{1}^{+-+}(\tau_{1},\tau_{2},\tau_{3})={}&\mathcal{E}^{+}_{1}\mathcal{E}^{-}_{2}\mathcal{E}^{+}_{3}\mathrm{Tr}[\hat{\mu}_{4}\hat{\mu}_{3}^+\hat{\mu}_{2}^-\hat{\mu}_{1}^+\hat{\rho}_{0}]\\
F_{2}^{+-+}(\tau_{1},\tau_{2},\tau_{3})={}&-\mathcal{E}^{+}_{1}\mathcal{E}^{-}_{2}\mathcal{E}^{+}_{3}\mathrm{Tr}[\hat{\mu}_{4}\hat{\mu}_{3}^+\hat{\mu}_{1}^+\hat{\rho}_{0}\hat{\mu}_{2}^-]\\
F_{4}^{+-+}(\tau_{1},\tau_{2},\tau_{3})={}&\mathcal{E}^{+}_{1}\mathcal{E}^{-}_{2}\mathcal{E}^{+}_{3}\mathrm{Tr}[\hat{\mu}_{4}\hat{\mu}_{1}^+\hat{\rho}_{0}\hat{\mu}_{2}^-\hat{\mu}_{3}^+]\,,
\end{align}
\end{subequations}
where here, the analytic signal is given by only the nonconjugate terms in Eq (\ref{eq:resp}). Note that there is no $F_{3}^{+-+}(\tau_{1},\tau_{2},\tau_{3})$ term. The explanation for why this term is zero is as follows: the first and third pulses are $B$ and $C$, which carry the same sign on their wave vectors $\pm\mathbf{k}_{\mathrm{B}}$ and $\pm\mathbf{k}_{\mathrm{C}}$. We know that the sign on pulse $B$ is positive, therefore the sign on pulse $C$ must also be positive. Due to the RWA,  a positive wavevector on the third pulse corresponds to a positive frequency component of the transition dipole moment---i.e. $\mathcal{E}_{3}^{+}\hat{\mu}_{3}^{+}$---and since in this term the third transition dipole moment acts on the ground state from the right, the entire term goes to zero.\\

The above pulse ordering does not lead to a photon echo.  Instead it leads to a free polarization decay (analogous to the well-known free induction decay (FID) in NMR spectroscopy) and therefore it is known as a \emph{nonrephasing} experiment. The last possibility is if pulse $A$ arrives third, e.g. ${T}_{B}<{T}_{C}<{T}_{A}$. In this case the only terms to survive the phase matching condition are
\begin{subequations}\label{eq:resp2q}
\begin{align}
F_{1}^{++-}(\tau_{1},\tau_{2},\tau_{3})={}&\mathcal{E}^{+}_{1}\mathcal{E}^{+}_{2}\mathcal{E}^{-}_{3}\mathrm{Tr}[\hat{\mu}_{4}\hat{\mu}_{3}^-\hat{\mu}_{2}^+\hat{\mu}_{1}^+\hat{\rho}_{0}]\\
F_{3}^{++-}(\tau_{1},\tau_{2},\tau_{3})={}&-\mathcal{E}^{+}_{1}\mathcal{E}^{+}_{2}\mathcal{E}^{-}_{3}\mathrm{Tr}[\hat{\mu}_{4}\hat{\mu}_{2}^+\hat{\mu}_{1}^+\hat{\rho}_{0}\hat{\mu}_{3}^-]\,.
\end{align}
\end{subequations}
where here, the analytic signal is also given by only the nonconjugate terms in Eq (\ref{eq:resp}). The above pulse ordering is known as a \emph{two-quantum} experiment, and it is used far less often than the rephasing and nonrephasing pulse orderings \cite{Kim2009,Kim2009a,Stone:2009rt,Stone:2009aa,Karaiskaj:2010aa,Turner:2010aa,Nemeth:2010aa,Dai:2012aa}. A similar line of reasoning to above can be used to see why terms $F_{2}^{++-}(\tau_{1},\tau_{2},\tau_{4})$ and $F_{1}^{++-}(\tau_{1},\tau_{2},\tau_{3})$ go to zero.

\subsection{Phenomenological treatment of dephasing and population relaxation}\label{sec:deph}

Non-unitary processes such as dephasing and population relaxation can be included phenomenologically; this is typically done in Liouville space using the Green's function formalism \cite{Mukamel:1995aa}. Here, we present an analogous treatment in Hilbert space which captures the essential effects of dephasing and population relaxation.\\

We begin by rewriting Eqs (\ref{eq:respJ}) using the definition for the time-dependent transition dipole moment operator given in Eq. (\ref{eq:mut}).
We then simplify this expression by making use of the unitary operator properties $\hat{U}({t}_{1}+{t}_{2}+\dots)=\hat{U}({t}_{1})\hat{U}({t}_{2})\dots$ and $\hat{U}\dg({t})\hat{U}({t})=I$ as well as the cyclic properties of the trace. This reduces to

\begin{subequations}\label{eq:respJ2}
\begin{align}\label{eq:resp7}
\begin{split}
F_{1}&^{{p}_{1}{p}_{2}{p}_{3}}(\tau_{1},\tau_{2},\tau_{3})={}\mathcal{E}^{{p}_{1}}_{1}\mathcal{E}^{{p}_{2}}_{2}\mathcal{E}^{{p}_{3}}_{3}\mathrm{Tr}[\hat{\mu}\hat{U}_{3}\hat{\mu}^{{p}_{3}}\hat{U}_{2}\hat{\mu}^{{p}_{2}}\hat{U}_{1}\hat{\mu}^{{p}_{1}}\rho_{0,\textsc{s}}\hat{U}\dg_{1}\hat{U}\dg_{2}\hat{U}\dg_{3}]
\end{split}\\
\begin{split}
F_{2}&^{{p}_{1}{p}_{2}{p}_{3}}(\tau_{1},\tau_{2},\tau_{3})={}-\mathcal{E}^{{p}_{1}}_{1}\mathcal{E}^{{p}_{2}}_{2}\mathcal{E}^{{p}_{3}}_{3}\mathrm{Tr}[\hat{\mu}\hat{U}_{3}\hat{\mu}^{{p}_{3}}\hat{U}_{2}\hat{U}_{1}\hat{\mu}^{{p}_{1}}\rho_{0,\textsc{s}}\hat{U}\dg_{1}\hat{\mu}^{{p}_{2}}\hat{U}\dg_{2}\hat{U}\dg_{3}]
\end{split}\\
\begin{split}
F_{3}&^{{p}_{1}{p}_{2}{p}_{3}}(\tau_{1},\tau_{2},\tau_{3})={}-\mathcal{E}^{{p}_{1}}_{1}\mathcal{E}^{{p}_{2}}_{2}\mathcal{E}^{{p}_{3}}_{3}\mathrm{Tr}[\hat{\mu}\hat{U}_{3}\hat{U}_{2}\hat{\mu}^{{p}_{2}}\hat{U}_{1}\hat{\mu}^{{p}_{1}}\rho_{0,\textsc{s}}\hat{U}\dg_{1}\hat{U}\dg_{2}\hat{\mu}^{{p}_{3}}\hat{U}\dg_{3}]
\end{split}\\
\begin{split}
F_{4}&^{{p}_{1}{p}_{2}{p}_{3}}(\tau_{1},\tau_{2},\tau_{3})={}\mathcal{E}^{{p}_{1}}_{1}\mathcal{E}^{{p}_{2}}_{2}\mathcal{E}^{{p}_{3}}_{3}\mathrm{Tr}[\hat{\mu}\hat{U}_{3}\hat{U}_{2}\hat{U}_{1}\hat{\mu}^{{p}_{1}}\rho_{0,\textsc{s}}\hat{U}\dg_{1}\hat{\mu}^{{p}_{2}}\hat{U}\dg_{2}\hat{\mu}^{{p}_{3}}\hat{U}\dg_{3}]\,,
\end{split}
\end{align}
\end{subequations}
where   $\rho_{0,\textsc{s}}=\hat{U}\dg({t})\hat{\rho}_{0}\hat{U}(t)$ is the initial state in the Schr\"odinger picture, $\hat{U}_{j}=\hat{U}(\tau_{j})$, and where $\hat{U}(t)=\ee{-\tfrac{i}{\hbar}\ham_{\textsc{s}}t}$. To include the particular non-unitary processes of interest, we then make an ad-hoc substitution $\hat{U}_{j}\rightarrow \hat{\Lambda}_{j}$ where
\begin{align}
\hat{\Lambda}_{j}={}&\ee{-\tfrac{i}{\hbar}\ham'_{\textsc{s}}\tau_{j}}\,,
\end{align}
and $\ham'_{\textsc{s}}$ is an unphysical `Hamiltonian' that corresponds to the system Hamiltonian $\ham_{\textsc{s}}$, with energies $\hbar\omega_{a}$ replaced by $\hbar(\omega_{a}+i\gamma_{a})$. We note that at this stage, $\gamma_{i}$ does not have a physical interpretation itself. Eqs. (\ref{eq:respJ2}) become  
\begin{subequations}\label{eq:respJ3}
\begin{align}\label{eq:resp7}
\begin{split}
F_{1}&^{{p}_{1}{p}_{2}{p}_{3}}(\tau_{1},\tau_{2},\tau_{3})={}\mathcal{E}^{{p}_{1}}_{1}\mathcal{E}^{{p}_{2}}_{2}\mathcal{E}^{{p}_{3}}_{3}\mathrm{Tr}[\hat{\mu}\hat\Lambda_{3}\hat{\mu}^{{p}_{3}}\hat\Lambda_{2}\hat{\mu}^{{p}_{2}}\hat\Lambda_{1}\hat{\mu}^{{p}_{1}}\rho_{0,\textsc{s}}\hat\Lambda\dg_{1}\hat\Lambda\dg_{2}\hat\Lambda\dg_{3}]
\end{split}\\\label{eq:resp7d}
\begin{split}
F_{2}&^{{p}_{1}{p}_{2}{p}_{3}}(\tau_{1},\tau_{2},\tau_{3})={}-\mathcal{E}^{{p}_{1}}_{1}\mathcal{E}^{{p}_{2}}_{2}\mathcal{E}^{{p}_{3}}_{3}\mathrm{Tr}[\hat{\mu}\hat\Lambda_{3}\hat{\mu}^{{p}_{3}}\hat\Lambda_{2}\hat\Lambda_{1}\hat{\mu}^{{p}_{1}}\rho_{0,\textsc{s}}\hat\Lambda\dg_{1}\hat{\mu}^{{p}_{2}}\hat\Lambda\dg_{2}\hat\Lambda\dg_{3}]
\end{split}\\
\begin{split}
F_{3}&^{{p}_{1}{p}_{2}{p}_{3}}(\tau_{1},\tau_{2},\tau_{3})={}-\mathcal{E}^{{p}_{1}}_{1}\mathcal{E}^{{p}_{2}}_{2}\mathcal{E}^{{p}_{3}}_{3}\mathrm{Tr}[\hat{\mu}\hat\Lambda_{3}\hat\Lambda_{2}\hat{\mu}^{{p}_{2}}\hat\Lambda_{1}\hat{\mu}^{{p}_{1}}\rho_{0,\textsc{s}}\hat\Lambda\dg_{1}\hat\Lambda\dg_{2}\hat{\mu}^{{p}_{3}}\hat\Lambda\dg_{3}]
\end{split}\\
\begin{split}
F_{4}&^{{p}_{1}{p}_{2}{p}_{3}}(\tau_{1},\tau_{2},\tau_{3})={}\mathcal{E}^{{p}_{1}}_{1}\mathcal{E}^{{p}_{2}}_{2}\mathcal{E}^{{p}_{3}}_{3}\mathrm{Tr}[\hat{\mu}\hat\Lambda_{3}\hat\Lambda_{2}\hat\Lambda_{1}\hat{\mu}^{{p}_{1}}\rho_{0,\textsc{s}}\hat\Lambda\dg_{1}\hat{\mu}^{{p}_{2}}\hat\Lambda\dg_{2}\hat{\mu}^{{p}_{3}}\hat\Lambda\dg_{3}]\,.
\end{split}
\end{align}
\end{subequations}

After evaluation of the expressions in Eqs. (\ref{eq:respJ3}), we make the substitution $(\gamma_{a}+\gamma_{b})\tau_{k}\rightarrow \Gamma_{ab}\tau_{k}$. This step can easily be automated with symbolic manipulation software such as \emph{Mathematica} \cite{Wolfram-Research2010}. We can then recognize $\Gamma_{aa}=1/T_{1}$ as the population relaxation rate for eigenstate $\ket{a}{}$, and $\Gamma_{ab}=1/T_{2}$ as the dephasing rate between eigenstates $\ket{a}{}$ and $\ket{b}{}$. \\

We note that in order to recover a physical picture, we could only perform the substitution $\hat{U}_{j}\rightarrow \hat{\Lambda}_{j}$ once the expressions were reduced to the form given by Eqs. (\ref{eq:respJ2}) and not earlier.  Eqs. (\ref{eq:respJ3}) are general and can now be used to treat any arbitrary system.

\section{Example: A heterodimer}
\label{sec:example}
In this section we demonstrate how to compute a 2D spectrum for a pair of two-level systems, i.e. qubits, a canonical example in quantum optics and spectroscopy and therefore an excellent candidate to study. The system is simple enough to treat analytically but complex enough to reveal the consequences of coupling.

\subsection{System description}\label{sec:system}

Each of the individual systems $a$ and $b$ consists of an electronic ground $\ket{{g}_{j}}{}$ and excited $\ket{{e}_{j}}{}$ state, where $j=a,b$. Assuming that orbital overlap between $a$ and $b$ is negligible \cite{Scholes1994}, then the total Hamiltonian for the material system is given by the Hamiltonians of the individual systems as well as the coupling Hamiltonian:
\begin{align}\label{eq:totham}
\ham_{\textsc{s}}={}&\hbar\omega_{a}\ket{{e}_{a}}{}\bra{{e}_{a}}{}+\hbar\omega_{b}\ket{{e}_{b}}{}\bra{{e}_{b}}{}+J\big(\hat{\sigma}^+_{a}\hat{\sigma}^-_{b}+\hc\big)\,,
\end{align}
where $\hbar\omega_{a/b}$ are the excited-state energies of the two individual systems, $J$ is the coupling energy between the two systems, and $\hat{\sigma}^+_{j}=\ket{{e}_{j}}{}\bra{{g}_{j}}{}$ and $\hat{\sigma}^-_{j}=\ket{{g}_{j}}{}\bra{{e}_{j}}{}$. We have set the ground-state energies to zero and $\omega_{b}\geq\omega_{a}$. The transition dipole moment operator is given by
\begin{align}
\hat{\mu}_{j}={\mu}_{j}\big(\ket{{e}_{j}}{}\bra{{g}_{j}}{}+\hc\big)\,,
\end{align}
where $j=a,b$ and we have set the transition dipoles to be real for simplicity. 

\begin{figure}
\begin{center}
\includegraphics[width=0.4\textwidth]{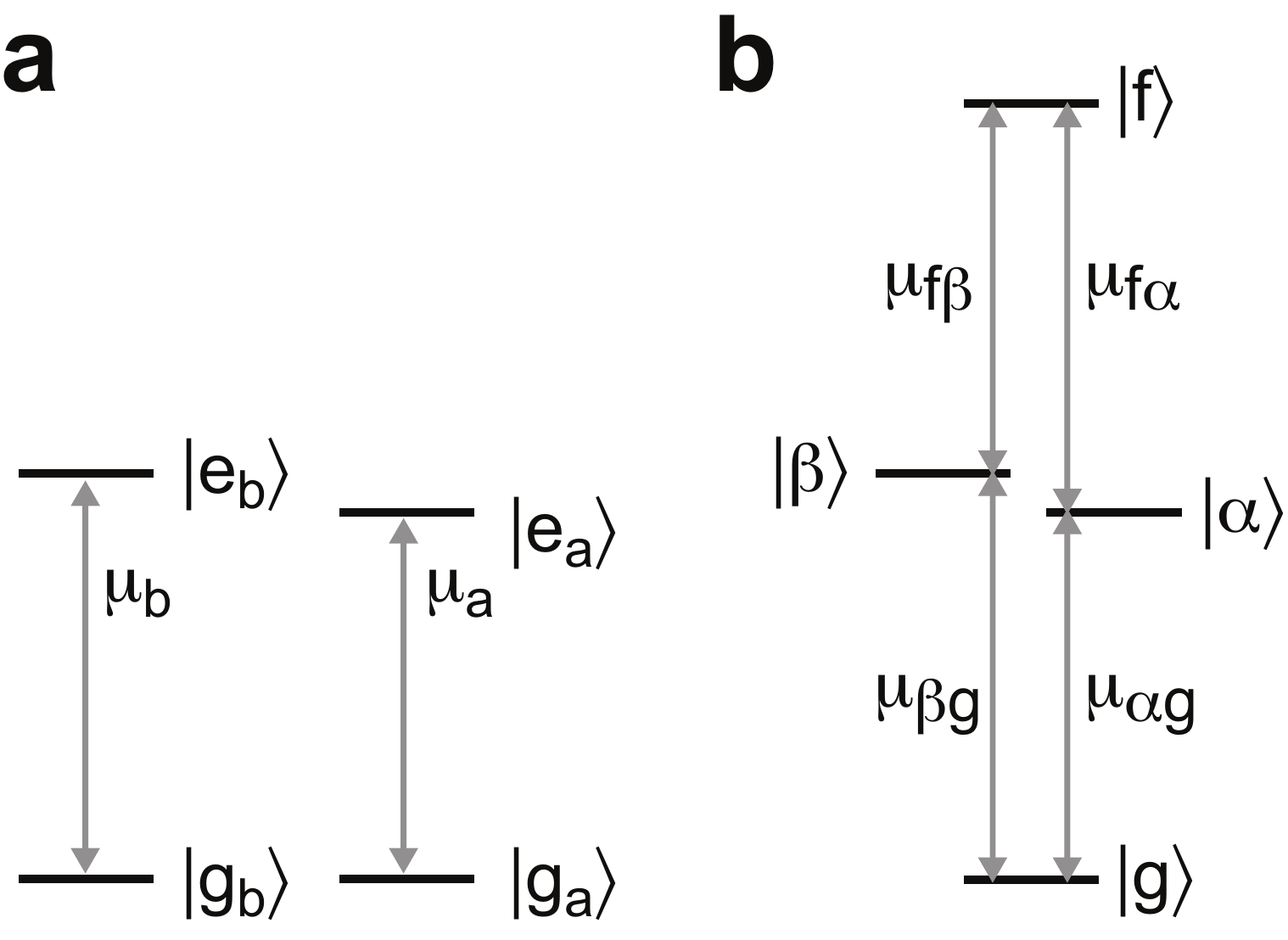}
\caption{Schematics of the energy-level structure for a pair of two-level systems, $a$ and $b$. View (a) depicts the heterodimer in the chromophore representation, where systems $a$ and $b$ occupy different Hilbert spaces. View (b) depicts the heterodimer as a composite system, as defined in Eq. (\ref{eq:states}). In the absence of coupling, all of the wave functions in view (b) can be written as product states, for example, $\vert \alpha \rangle = \vert e_a \rangle \vert g_b \rangle$.} 
\end{center}
\label{fig:elevels}
\end{figure}

In the standard nomenclature of the field, the eigenbasis of the individual systems $a$ and $b$ is known as the \emph{site} basis, whereas the eigenbasis of the total material system Hamiltonian is known as the \emph{exciton} basis. Diagonalization of Eq.  (\ref{eq:totham})  allows the total Hamiltonian to be written in the exciton basis:
\begin{align}\label{eq:totham7}
\ham_{\textsc{s}}={}&\hbar\omega_{\alpha}\ket{\alpha}{}\bra{\alpha}{}+\hbar\omega_{\beta}\ket{\beta}{}\bra{\beta}{}+\hbar\omega_{f}\ket{f}{}\bra{f}{}\,,
\end{align}
where $\omega_{f}=\omega_{a}+\omega_{b}$, and  the single-exciton energies
\begin{align}
\omega_{\alpha/\beta}=\omega\pm\Delta\mathrm{sec}(2\theta)
\end{align}
 are expressed in terms of the following convenient parameters: the
average of the site energies $\omega=(\omega_{a}+\omega_{b})/2$, the difference $\Delta=(\omega_{a}-\omega_{b})/2$, and the mixing angle $\theta=\mathrm{arctan}(J/\hbar\Delta)/2$. The exciton basis is given by
 \begin{subequations}\label{eq:states}
 \begin{align}
 \ket{g}{}={}&\ket{{g}_{a}}{}\ket{{g}_{b}}{}\\
 \begin{split}
 \ket{{\alpha}}{}={}&\cos(\theta)\ket{{e}_{a}}{}\ket{{g}_{b}}{}+\sin(\theta)\ket{{g}_{a}}{}\ket{{e}_{b}}{}
 \end{split}\\
 \begin{split}
 \ket{{\beta}}{}={}&-\sin(\theta)\ket{{e}_{a}}{}\ket{{g}_{b}}{}+\cos(\theta)\ket{{g}_{a}}{}\ket{{e}_{b}}{}
 \end{split}\\
  \ket{f}{}={}&\ket{{e}_{a}}{}\ket{{e}_{b}}{}\,.
 \end{align}
 \end{subequations}
The transition dipole moment operator for the entire system can also be written in the exciton basis
\begin{align}
\begin{split}
\hat{\mu}={}&\mu_{\alpha g}\ket{{\alpha}}{}\bra{{g}}{}+\mu_{\beta g}\ket{{\beta}}{}\bra{{g}}{}+\mu_{f\alpha}\ket{{f}}{}\bra{{\alpha}}{}+\mu_{f\beta}\ket{{f}}{}\bra{{\beta}}{}+\hc ,
\end{split}
\end{align}
where
\begin{align}
\left[\begin{array}{c}\mu_{\alpha g} \\\mu_{\beta g}\end{array}\right]={}&\left[\begin{array}{cc}\cos (\theta) & \sin(\theta) \\ -\sin(\theta) & \cos (\theta)\end{array}\right]\left[\begin{array}{c}\mu_{a} \\\mu_{b}\end{array}\right]\\
\left[\begin{array}{c}\mu_{f\alpha} \\\mu_{f\beta}\end{array}\right]={}&\left[\begin{array}{cc}\sin (\theta) & \cos(\theta) \\ \cos(\theta) & -\sin (\theta)\end{array}\right]\left[\begin{array}{c}\mu_{a} \\\mu_{b}\end{array}\right]\,.
\end{align}

Notice that when there is no coupling, i.e. $J=0$ and therefore $\theta=0$, the transition dipole moments reduce to $\mu_{\alpha g}=\mu_{f\beta}=\mu_{a}$ and $\mu_{\beta g}=\mu_{f\alpha}=\mu_{b}$. The consequences of this will be discussed in the next section.\\

To predict the spectra, we require the transition dipole moment operator in the interaction picture, given by Eq. (\ref{eq:mut}), which we express in terms of the positive and negative frequency components:
\begin{align}\label{eq:muthing}
\hat{\mu}_{\textsc{i}}(t)={}&\hat{\mu}^+_{\textsc{i}}(t)+\hat{\mu}^-_{\textsc{i}}(t)
\end{align}
where $\hat{\mu}^+_{\textsc{i}}(t)={}\big(\hat{\mu}^-_{\textsc{i}}(t)\big)\dg$ and
\begin{align}\label{eq:tdo}
\begin{split}
\hat{\mu}^+_{\textsc{i}}(t)={}&\hat{\mu}_{\alpha g}\ee{+i\omega_{\alpha}{t}}\ket{{\alpha}}{}\bra{{g}}{}+\hat{\mu}_{\beta g}\ee{+i\omega_{\beta}{t}}\ket{{\beta}}{}\bra{{g}}{}+\hat{\mu}_{f\alpha}\ee{+i\omega_{\beta}{t}}\ket{{f}}{}\bra{{\alpha}}{}+\hat{\mu}_{f\beta}\ee{+i\omega_{\alpha}{t}}\ket{{f}}{}\bra{{\beta}}{}
\end{split}
\end{align}
where we assumed $\omega_{g}=0$ and made use of $\omega_{f}-\omega_{\alpha}=\omega_{\beta}$ and $\omega_{f}-\omega_{\beta}=\omega_{\alpha}$.

\subsection{Predict 2D spectra}
Recall that one of the advantages of 2D ES over conventional transient-absorption spectroscopy is the ability to select different parts of the total signal by controlling the properties of the incoming pulses, most often the relative time ordering. This is afforded by the noncollinear beam geometry which makes use of phase matching. In this section, we will use the formalism developed in Sec. \ref{sec:theory} to predict rephasing, non-rephasing and two-quantum spectra for  the heterodimer system described above. Here, we disregard non-unitary processes and focus on 2D spectral properties such as location, height, and oscillatory behaviour of peaks. We will extend our example to include dephasing and population relaxation in the next section.\\

We first consider a rephasing experiment, where pulse $A$ arrives first, ${T}_{\mathrm{A}}<{T}_{\mathrm{B}}<{T}_{\mathrm{C}}$. The only terms in the expression for the third-order polarization in Eq. (\ref{eq:resp}) are those given in Eq. (\ref{eq:respR}). Using the above transition dipole moment operator and evaluating the trace functions yields the following expression for the analytic signal
\begin{align}
\begin{split}
E_{\mathrm{R}}(\tau_{1},\tau_{2},\tau_{3})\propto{}&\mathcal{E}^-_{A}\mathcal{E}^+_{B} \mathcal{E}^+_{C}{}\theta(\tau_{1}) \theta(\tau_{2}) \theta(\tau_{3})\Big(2 \mu _{\alpha {g}}^4 \ee{+i\omega_{\alpha}\tau_{1}}\ee{-i\omega_{\alpha}\tau_{3}}+2 \mu _{\beta {g}}^4 \ee{+i\omega_{\beta}\tau_{1}}\ee{-i\omega_{\beta}\tau_{3}}\\
&+   \Big(\mu _{\beta {g}}^2 \big(\mu _{\alpha {g}}^2-\mu _{{f} \beta}^2\big)+\mu _{\alpha {g}}\mu _{\beta {g}}  \big(\mu _{\alpha {g}} \mu _{\beta {g}}-\mu _{{f} \alpha} \mu _{{f} \beta}\big)e^{+i  \omega _{\beta \alpha }\tau _2}\Big)\ee{+i\omega_{\beta}\tau_{1}}\ee{-i\omega_{\alpha}\tau_{3}} \\
&+  \Big(\mu _{\alpha {g}}^2 \big(\mu _{\beta {g}}^2-\mu _{{f} \alpha}^2\big)+\mu _{\alpha {g}}\mu _{\beta {g}} \big(\mu _{\alpha {g}} \mu _{\beta {g}}-\mu _{{f} \alpha} \mu _{{f} \beta}\big)e^{-i  \omega _{\beta \alpha }\tau _2} \Big)\ee{+i\omega_{\alpha}\tau_{1}}\ee{-i\omega_{\beta}\tau_{3}} \Big)\,.
\end{split}
\end{align}

Following convention, we Fourier transform $E_{\mathrm{R}}(\tau_{1},\tau_{2},\tau_{3})$ with respect to the $\tau_1$ and $\tau_3$ variables.
\begin{align}
\begin{split}
E_{\mathrm{R}}^{(3)}&(\omega_{1},\tau_{2},\omega_{3})={}\iint d\tau_{1}\tau_{3}E_{\mathrm{R}}^{(3)}(\tau_{1},\tau_{2},\tau_{3})e^{+i\omega_{1}\tau_{1}}e^{+i\omega_{3}\tau_{3}}\, .
\end{split}
\end{align}

For the sake of illustration, we temporarily disregard the Heaviside step functions $\theta(\tau_{1}) \theta(\tau_{2}) \theta(\tau_{3})$, resuming a complete treatment in the next section on dephasing and population relaxation. The Fourier transform is then

\begin{align}\label{eq:repheq}
\begin{split}
E_{\mathrm{R}}^{(3)}(\omega_{1},\tau_{2},\omega_{3})\propto{}&\mathcal{E}^-_{A}\mathcal{E}^+_{B} \mathcal{E}^+_{C}{}\Bigg[2 \mu _{\alpha {g}}^4 \delta (\omega_{1}+\omega _{\alpha }) \delta (\omega_{3}-\omega _{\alpha })+2 \mu _{\beta {g}}^4 \delta (\omega_{1}+\omega _{\beta }) \delta (\omega_{3}-\omega _{\beta })\\ 
&+  \Big(\mu _{\alpha {g}}^2 \big(\mu _{\beta {g}}^2-\mu _{{f} \alpha}^2\big)+\mu _{\alpha {g}}\mu _{\beta {g}} \big(\mu _{\alpha {g}} \mu _{\beta {g}}-\mu _{{f} \alpha} \mu _{{f} \beta}\big)e^{-i  \omega _{\beta \alpha }\tau _2} \Big)\delta (\omega_{1}+\omega _{\alpha }) \delta (\omega_{3}-\omega _{\beta })\\ 
&+  \Big(\mu _{\beta {g}}^2 \big(\mu _{\alpha {g}}^2-\mu _{{f} \beta}^2\big)+\mu _{\alpha {g}}\mu _{\beta {g}}  \big(\mu _{\alpha {g}} \mu _{\beta {g}}-\mu _{{f} \alpha} \mu _{{f} \beta}\big)e^{+i  \omega _{\beta \alpha }\tau _2}\Big)\delta (\omega_{1}+\omega _{\beta })\delta (\omega_{3}-\omega _{\alpha })\Bigg]\,.
\end{split}
\end{align}

\begin{figure*}[t!]
\centering
  \includegraphics[width=0.8\textwidth]{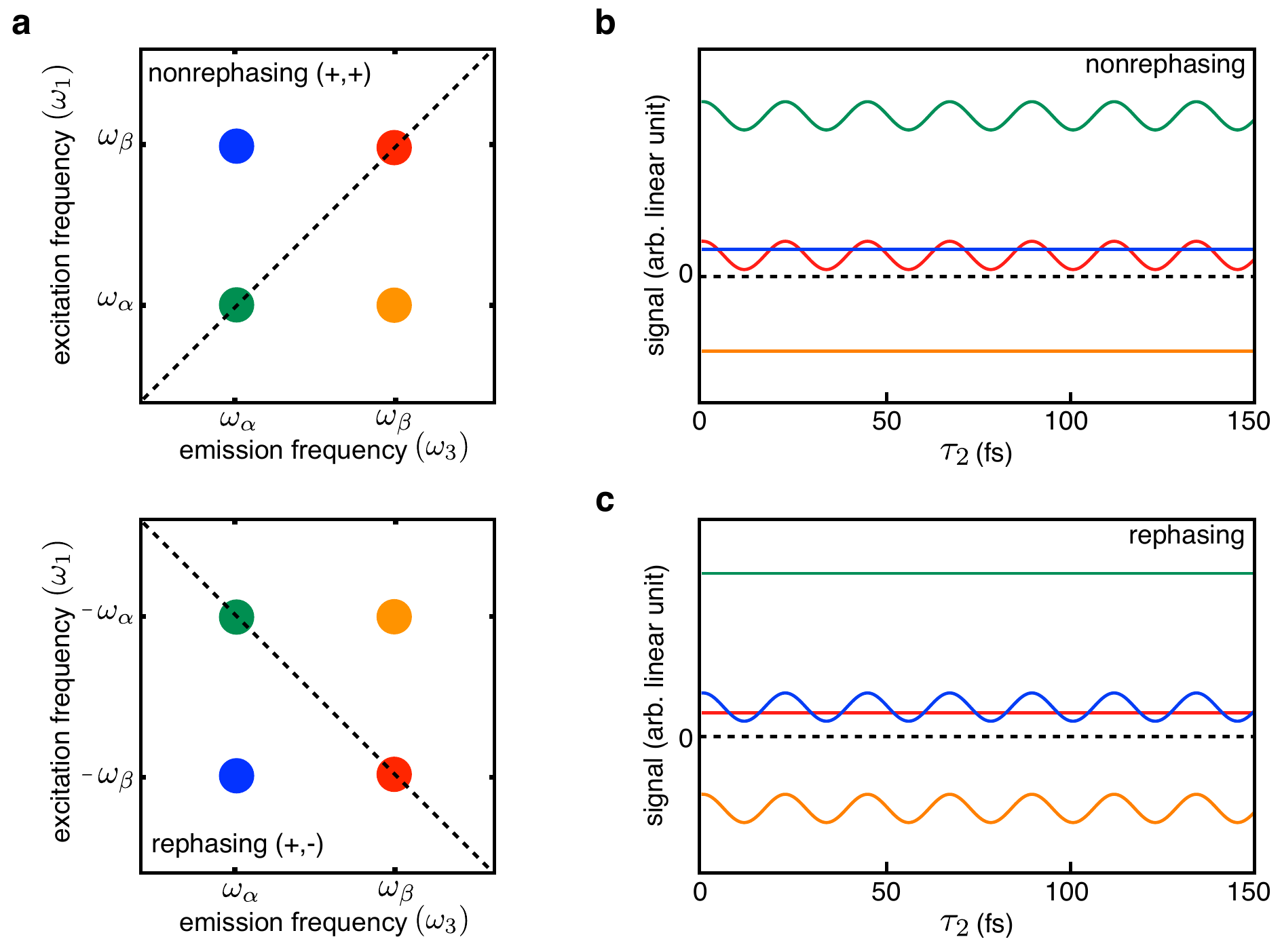}
  \caption{2D ES for a heterodimer undergoing unitary dynamics. (a) 2D spectrum for rephasing ($E_{\mathrm{R}}^{(3)}(\omega_{1},0,\omega_{3})$ in Eq. \ref{eq:repheq}) and nonrephasing ($E_{\mathrm{nR}}^{(3)}(\omega_{1},0,\omega_{3})$ in Eq. \ref{eq:nonrepheq}) pulse orderings. (b) and (c) Dynamics of peaks for rephasing and nonrephasing pulse orderings, respectively. In a rephasing (nonrephasing) spectrum the cross peaks (diagonal peaks) oscillate as a function of delay $\tau_2$. For a homodimer, the splitting between the peaks, and the oscillation frequency for oscillating peaks, are $2J$. The traces were \emph{not} vertically shifted, as is typically done with real data. Parameters used: $\nu_{a}=365 ~\mathrm{THz} ~(1510~ \mathrm{meV})$, $\nu_{b}=397~ \mathrm{THz} ~(1640~ \mathrm{meV})$, $J/ \hbar = 50 \pi^{-1}~ \mathrm{THz} ~(66~ \mathrm{meV})$, $\mu_{a}=-1.1~\mathrm{D}$,  $\mu_{b}=1.5~\mathrm{D}$ and  $E_{j}=1 ~\mathrm{N}/\mathrm{C}$ for $j=1,2,3$.}
  \label{fig:2dspec}
\end{figure*}

If we plot the above signal as a function of $\omega_1$ and $\omega_3$, we obtain four peaks, as shown in Fig. \ref{fig:2dspec}a. Their amplitudes are determined by the transition dipoles and we bring attention to the fact that the cross peaks oscillate as a function of $\tau_{2}$ at a frequency $\omega_{\beta\alpha}$ determined by the energy difference between the excited states $\ket{\alpha}{}$ and $\ket{\beta}{}$, as shown in Fig. \ref{fig:2dspec}b. At first glance, it may seem that the cross peaks should oscillate out-of-phase due to the $e^{-i  \omega _{\beta \alpha }\tau _2}$ and $e^{+i  \omega _{\beta \alpha }\tau _2}$ prefactors; however, recall that the real electric field is twice the real part of its analytic signal ${E}({t})$. The $\tau_{2}$ dependence of both cross peaks will therefore  have the same form since $\cos(\omega _{\beta \alpha }\tau _2)=\cos(-\omega _{\beta \alpha }\tau _2)$. Other phase relationships between peaks can also be predicted with a more elaborate treatment, such as the one considered by Butkus \emph{et al.} \cite{Butkus2012}.

We can follow the same procedure to produce 2D spectra for the nonrephasing (nR) experiment, where pulse $A$ arrives second, e.g. ${T}_{\mathrm{B}}<{T}_{\mathrm{A}}<{T}_{\mathrm{C}}$. The signal is given by
\begin{align}\label{eq:nonrepheq}
\begin{split}
E_{\textrm{nR}}^{(3)}(\omega_{1},\tau_{2},\omega_{3})\propto{}&\mathcal{E}^+_{B}\mathcal{E}^-_{A} \mathcal{E}^+_{C}{}\Bigg[\mu _{\alpha {g}}^2 \big(\mu _{\beta {g}}^2-\mu _{{f} \alpha}^2\big) \delta (\omega_{1}-\omega _{\alpha }) \delta (\omega_{3}-\omega _{\beta })+\mu _{\beta {g}}^2 \big(\mu _{\alpha {g}}^2-\mu _{{f} \beta}^2\big)  \delta (\omega_{1}-\omega _{\beta })\delta (\omega_{3}-\omega _{\alpha })\\ 
&+ \Big(2 \mu _{\alpha {g}}^4+\mu _{\alpha {g}} \mu _{\beta {g}}  \big(\mu _{\alpha {g}} \mu _{\beta {g}}-\mu _{{f} \alpha} \mu _{{f} \beta}\big)e^{+i  \omega _{\beta \alpha }\tau _2}\Big)\delta (\omega_{1}-\omega _{\alpha }) \delta (\omega_{3}-\omega _{\alpha })\\ 
&+\Big(2 \mu _{\beta {g}}^4+\mu _{\alpha {g}} \mu _{\beta {g}}  \big(\mu _{\alpha {g}} \mu _{\beta {g}}-\mu _{{f} \alpha} \mu _{{f} \beta}\big)e^{-i  \omega _{\beta \alpha }\tau _2}\Big)\delta (\omega_{1}-\omega _{\beta }) \delta (\omega_{3}-\omega _{\beta }) \Bigg] \,.
\end{split}
\end{align}
Notice that in a nonrephasing experiment, it is the diagonal peaks that oscillate at the difference frequency during time period $\tau_2$, see Fig. \ref{fig:2dspec}d. The diagonal peaks also oscillate in phase with each other.\\

2D spectra are usually displayed as the sum of the rephasing and nonrephasing signals, where the rephasing signal is flipped into the $(+,+)$ quadrant. This eliminates a problem known as phase twist \cite{Ernst:1987aa} that is endemic to half-Fourier transforms. Producing a full-Fourier transform in this manner sharpens lineshapes and has certain advantages for information extraction \cite{Khalil:2003aa,Khalil:2003ab}. \\

The third option is the two-quantum (2Q) experiment, where pulse $A$ arrives third, e.g. ${T}_{\mathrm{B}}<{T}_{\mathrm{C}}<{T}_{\mathrm{A}}$. This pulse sequence is often used to isolate signals involving the doubly-excited state during the second time period, $\tau_2$. Such signals are most naturally expressed as a 2Q spectrum correlating the two-quantum frequencies to the emission frequencies. Therefore the Fourier transformation proceeds as
\begin{align}
E_{\textrm{2Q}}^{(3)}(\tau_{1},\omega_{2},\omega_{3})={}&\iint d\tau_{2}\tau_{3}E_{\textrm{2Q}}^{(3)}(\tau_{1},\tau_{2},\tau_{3})e^{+i\omega_{2}\tau_{2}}e^{+i\omega_{3}\tau_{3}}\,,
\end{align}
which yields
\begin{align}
\begin{split}
E_{\textrm{2Q}}^{(3)}(0,\omega_{2},\omega_{3})\propto{} &\mathcal{E}^+_{B}{}\mathcal{E}^+_{C}\mathcal{E}^-_{A}\Big[\big(\mu _{\alpha {g}} \mu _{{f} \alpha} +\mu _{\beta {g}} \mu _{{f} \beta}\big) \big(\mu _{\alpha {g}} \mu _{{f} \alpha}-\mu _{\beta {g}} \mu _{{f} \beta}\big) \delta (\omega_{3}-\omega _{\alpha })\\ 
&+\big(\mu _{\beta {g}} \mu _{{f} \beta}+\mu _{\alpha {g}} \mu _{{f} \alpha} \big)  \big(\mu _{\beta {g}} \mu _{{f} \beta}-\mu _{\alpha {g}} \mu _{{f} \alpha}\big) \delta (\omega_{3}-\omega _{\beta })\Big]\delta(\omega_{2}-\omega_{f})\,,
\end{split}
\end{align}
where all peaks oscillate at the frequency corresponding to the doubly-excited state during $\tau_2$. In most cases researchers are interested in the 2D spectrum for $\tau_1 = 0$. \\

In the case of no coupling, the transition dipole moments reduce to $\mu_{\alpha g}=\mu_{f\beta}=\mu_{a}$ and $\mu_{\beta g}=\mu_{f\alpha}=\mu_{b}$.  Certain terms then go to zero and the above signals reduce to
\begin{subequations}
\begin{align}
\begin{split}
E_{\mathrm{R}}^{(3)}(\omega_{1},\tau_{2},\omega_{3})\propto{}&\mathcal{E}^-_{A}\mathcal{E}^+_{B} \mathcal{E}^+_{C}{}\Big[ \mu _{a}^4 \delta (\omega_{1}+\omega _{\alpha }) \delta (\omega_{3}-\omega _{\alpha })+ \mu _{b}^4 \delta (\omega_{1}+\omega _{\beta }) \delta (\omega_{3}-\omega _{\beta })\Big]\end{split}\\
\begin{split}
E_{\textrm{nR}}^{(3)}(\omega_{1},\tau_{2},\omega_{3})\propto{}&\mathcal{E}^+_{B}\mathcal{E}^-_{A} \mathcal{E}^+_{C}{}\Big[  \mu _{a}^4\delta (\omega_{1}-\omega _{\alpha }) \delta (\omega_{3}-\omega _{\alpha })+ \mu _{b}^4\delta (\omega_{1}-\omega _{\beta }) \delta (\omega_{3}-\omega _{\beta }) \Big]\,,
\end{split}
\end{align}
\end{subequations}
where we no longer obtain cross peaks. The presence of cross peaks in a rephasing or nonrephasing 2D spectrum is therefore a signature of coupling between two systems. For a two-quantum experiment, we observe no signal in the case of no coupling
\begin{align}
E_{\textrm{2Q}}^{(3)}(0,\omega_{2},\omega_{3})= {}&0\,. 
\end{align}
This is expected since individual two-level systems do not generate two-quantum signals.

\subsection{Dephasing and population relaxation}
The above treatment can be extended to include non-unitary processes, as described in Sec. \ref{sec:deph}. For a rephasing experiment, the only terms in the expression for the third-order polarization in Eq. (\ref{eq:resp}) are $F_{1}^{-++}(\tau_{1},\tau_{2},\tau_{3})^*$, $F_{2}^{-++}(\tau_{1},\tau_{2},\tau_{3})^*$, $F_{3}^{-++}(\tau_{1},\tau_{2},\tau_{3})^*$ and $F_{4}^{-++}(\tau_{1},\tau_{2},\tau_{3})^*$ where $F_{r}^{{p}_{1}{p}_{2}{p}_{3}}(\tau_{1},\tau_{2},\tau_{3})$ are defined in Eq. (\ref{eq:respJ3}). Using the transition dipole moment operator for a coupled dimer derived in Section \ref{sec:system} and evaluating the trace functions yields the following expression for the analytic signal
\begin{align}\label{eq:repheqx}
\begin{split}
E_{\mathrm{R}}(\omega_{1},\tau_{2},\omega_{3})\propto{}&\mathcal{E}^-_{A}\mathcal{E}^+_{B} \mathcal{E}^+_{C}{}\theta(\tau_{2})\Bigg[\mu _{\alpha g}^4 (e^{- i\Omega _{gg}\tau_{2}}+e^{- i\Omega _{\alpha \alpha }\tau_{2}} ) G\left(\omega_{1}+\Omega^*_{\alpha g}\right)G\left(\omega_{3}-\Omega _{\alpha g}\right)\\
&+\mu _{\beta g}^4 (e^{- i\Omega _{gg}\tau_{2}} +e^{- i\Omega _{\beta \beta }\tau_{2}})G\left(\omega_{1}+\Omega^*_{\beta g}\right)G\left(\omega_{3}-\Omega _{\beta g}\right) \\
& -\mu _{\alpha g}\mu _{f\alpha }( \mu _{\alpha g}\mu _{f\alpha }e^{- i\Omega _{\alpha \alpha }\tau_{2}}  +\mu _{\beta g} \mu _{f\beta } e^{-i  \Omega _{\beta \alpha }\tau_{2}} ) G\left(\omega_{1}+\Omega^*_{\alpha g}\right)G\left(\omega_{3}-\Omega _{f\alpha }\right)\\
&+\mu _{\alpha g}^2 \mu _{\beta g}^2 (e^{- i\Omega _{gg}\tau_{2}} + e^{-i  \Omega _{\beta \alpha }\tau_{2}})G\left(\omega_{1}+\Omega^*_{\alpha g}\right) G\left(\omega_{3}-\Omega _{\beta g}\right)\\
&-\mu _{\beta g} \mu _{f\beta }( \mu _{\beta g} \mu _{f\beta }e^{- i\Omega _{\beta \beta }\tau_{2}}+\mu _{\alpha g} \mu _{f\alpha }  e^{+i  \Omega^*_{\beta \alpha }\tau_{2}} ) G\left(\omega_{1}+\Omega^*_{\beta g}\right)G\left(\omega_{3}-\Omega _{f\beta }\right)\\
&+\mu _{\alpha g}^2 \mu _{\beta g}^2 (e^{- i\Omega _{gg}\tau_{2}}+ e^{+i  \Omega^*_{\beta \alpha }\tau_{2}} )G\left(\omega_{1}+\Omega^*_{\beta g}\right)G\left(\omega_{3}-\Omega _{\alpha g}\right) \Bigg]\,,
\end{split}
\end{align}
where $\Omega_{ij}=\omega_{i,j}-i\Gamma_{ij}$, $\omega_{ij}=\omega_{i}-\omega_{j}$ and $G(x)=\sqrt{2\pi}(\delta(x)+i/\pi x)$. If $\Gamma_{ii}=\Gamma_{ij}=0$, $\omega_{g}=0$ and $\omega_{f}=\omega_{\alpha}+\omega_{b}$, this reduces to the expression in Eq. (\ref{eq:repheq}), up to the Heaviside step function that was excluded in Eq. (\ref{eq:repheq}) for illustrative purposes.\\

For a nonrephasing experiment, the only terms in the expression for the third-order polarization in Eq. (\ref{eq:resp}) are $F_{1}^{+-+}(\tau_{1},\tau_{2},\tau_{3})$, $F_{2}^{+-+}(\tau_{1},\tau_{2},\tau_{3})$ and $F_{4}^{+-+}(\tau_{1},\tau_{2},\tau_{3})$, where $F_{r}^{{p}_{1}{p}_{2}{p}_{3}}(\tau_{1},\tau_{2},\tau_{3})$ are defined in Eq. (\ref{eq:respJ3}). Using the transition dipole moment operator for a coupled dimer derived in Section \ref{sec:system} and evaluating the trace functions yields the following expression for the analytic signal

\begin{align}\label{eq:nonrepheq2}
\begin{split}
E_{\textrm{nR}}^{(3)}(\omega_{1},\tau_{2},\omega_{3})\propto{}&\mathcal{E}^+_{B}\mathcal{E}^-_{A} \mathcal{E}^+_{C}{}\theta(\tau_{2})\Bigg[\mu _{\alpha  g}^2 (\mu _{\beta  g}^2 e^{-i  \Omega _{g g}\tau_{2}}-\mu _{f \alpha }^2  e^{-i  \Omega _{\alpha  \alpha }\tau_{2}}) G\left(\omega_{1}-\Omega _{\alpha  g}\right) G\left(\omega_{3}-\Omega _{\beta  g}\right)\\
&+\mu _{\beta  g}^2 (\mu _{\alpha  g}^2  e^{-i  \Omega _{g g}\tau_{2}} -\mu _{f \beta }^2 e^{-i  \Omega _{\beta  \beta }\tau_{2}} )G\left(\omega_{1}-\Omega _{\beta  g}\right) G\left(\omega_{3}-\Omega _{\alpha  g}\right)\\
&+\Big(\mu _{\alpha  g}^4 (e^{-i  \Omega _{g g}\tau_{2}} +e^{-i  \Omega _{\alpha  \alpha }\tau_{2}})+\mu^2 _{\alpha  g} \mu^2 _{\beta  g}e^{+i  \Omega^*_{\beta  \alpha }\tau_{2}} \Big) G\left(\omega_{1}-\Omega _{\alpha  g}\right) G\left(\omega_{3}-\Omega _{\alpha g }\right)\\
& -\mu _{\alpha  g} \mu _{\beta  g}\mu _{f \alpha } \mu _{f \beta }  e^{+i  \Omega^*_{\beta  \alpha }\tau_{2}} G\left(\omega_{1}-\Omega _{\alpha  g}\right) G\left(\omega_{3}-\Omega _{f \beta }\right) \\
&+\Big(\mu _{\beta  g}^4 (e^{-i  \Omega _{g g}\tau_{2}} +e^{-i  \Omega _{\beta  \beta }\tau_{2}})+\mu^2 _{\alpha  g} \mu^2 _{\beta  g}e^{-i  \Omega _{\beta  \alpha }\tau_{2}} \Big)G\left(\omega_{1}-\Omega _{\beta  g}\right) G\left(\omega_{3}-\Omega _{\beta g }\right)\\
& - \mu _{\alpha  g} \mu _{\beta  g}\mu _{f \alpha } \mu _{f \beta } e^{-i  \Omega _{\beta  \alpha }\tau_{2}}  G\left(\omega_{1}-\Omega _{\beta  g}\right) G\left(\omega_{3}-\Omega _{f \alpha }\right)\Bigg] \,.
\end{split}
\end{align}

\begin{figure*}[t!]
\centering
  \includegraphics[width=0.8\textwidth]{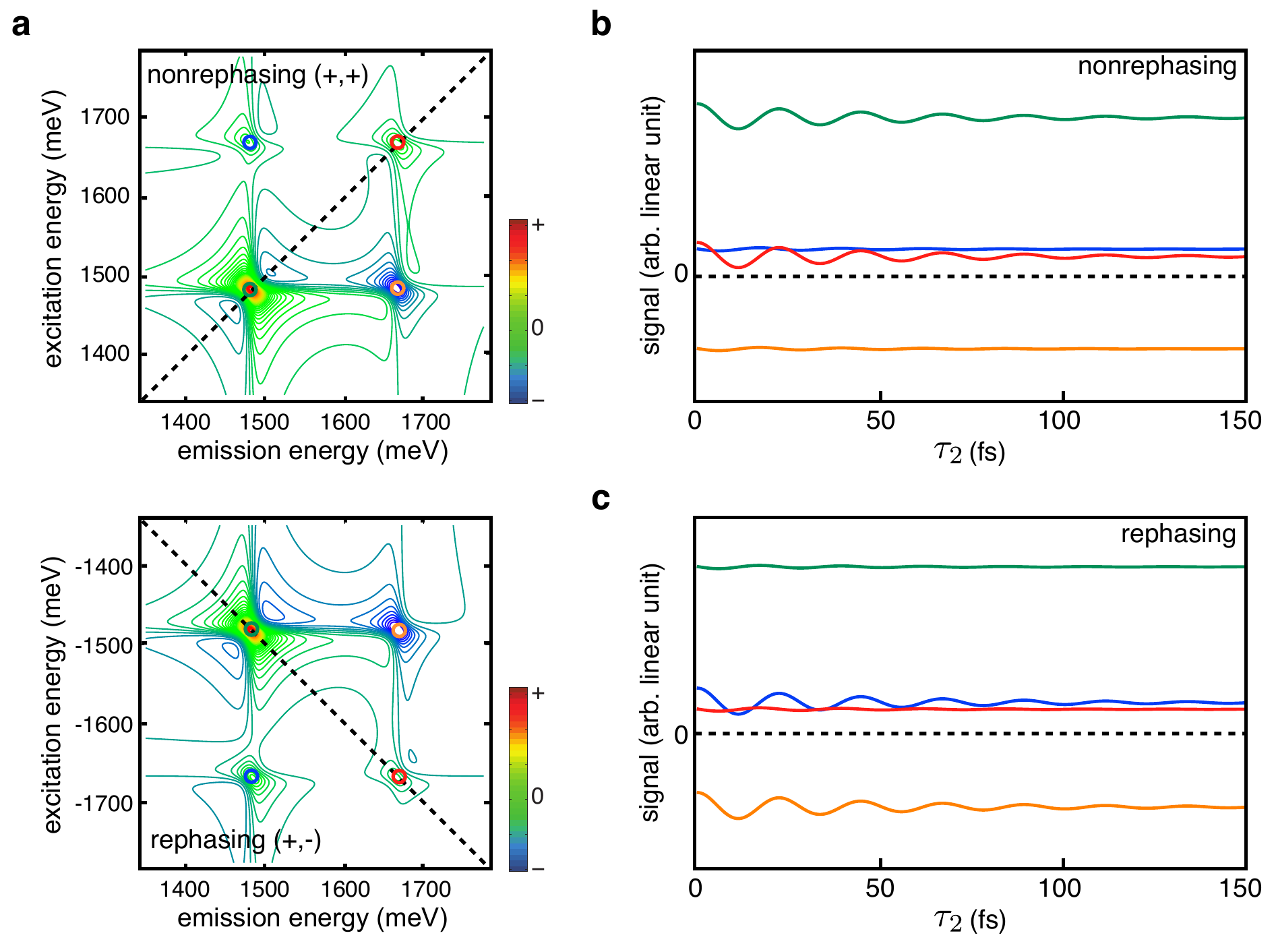}
  \caption{2D ES for a heterodimer undergoing non-unitary dynamics. (a) 2D spectrum for rephasing ($E_{\mathrm{R}}^{(3)}(\omega_{1},0,\omega_{3})$ in Eq. \ref{eq:repheq}) and nonrephasing ($E_{\mathrm{nR}}^{(3)}(\omega_{1},0,\omega_{3})$ in Eq. \ref{eq:nonrepheq}) pulse orderings. (b) and (c) Dynamics of peaks for rephasing and nonrephasing pulse orderings, respectively. In a rephasing (nonrephasing) spectrum the cross peaks (diagonal peaks) oscillate as a function of delay $\tau_2$. For a homodimer, the splitting between the peaks, and the oscillation frequency for oscillating peaks, are $2J$. The traces were \emph{not} vertically shifted, as is typically done with real data. Parameters used: $\nu_{a}=365 ~\mathrm{THz} ~(1510~ \mathrm{meV})$, $\nu_{b}=397~ \mathrm{THz} ~(1640~ \mathrm{meV})$, $J/ \hbar = 50 \pi^{-1}~ \mathrm{THz} ~(66~ \mathrm{meV})$, $\mu_{a}=-1.1~\mathrm{D}$,  $\mu_{b}=1.5~\mathrm{D}$,  $E_{j}=1~ \mathrm{N}/\mathrm{C}$ for $j=1,2,3$, $\Gamma_{ii}=0~\forall~i$ and $\Gamma_{ij}=0.02~\mathrm{fs}^{-1}~\forall~i,j$.}
  \label{fig:spec2}
\end{figure*}

Fig. \ref{fig:spec2} shows Eqs. (\ref{eq:repheqx}) and (\ref{eq:nonrepheq2}) for the parameters used in Fig.  \ref{fig:2dspec} and $\Gamma_{ii}=0~\forall~i$ and $\Gamma_{ij}=0.02\times 10^{15}~\mathrm{s}^{-1}~\forall~i,j$. Notice that the peaks now take on a finite width, where the spectral broadness is determined by the dephasing parameters. If we examine the peaks and cross-peaks at different waiting times $\tau_{2}$, we see that the oscillations decay on a time scale $T_{2}=1/\Gamma_{ij}=50 ~\mathrm{fs}$. Slight oscillations in diagonal peaks (cross peaks) for rephasing (nonrephasing) spectra result from their overlap with the strongly oscillating cross peaks (diagonal peaks).\\

\subsection{Measurements of GaAs quantum wells}
GaAs quantum wells have been studied for decades.  Several reviews detail how each advance in spectroscopic methodology has led to new insights into the excitonic many-body interactions, and 2D ES has been no exception \cite{Shah:1999aa,Cundiff:2008aa,Cundiff:2012aa}. For our purposes, the two lowest-energy exciton states of a GaAs quantum well sample can serve as an example of a heterodimer. The heavy-hole (excitation at 1540 meV) and light-hole exciton (excitation at 1546 meV) states are coupled because they involve electrons that occupy the same conduction band.\\

One important distinction between this sample and the model above is that the double-excitation states---the biexcitons---are energetically red-shifted from the sum of the single-exciton energies by about 1.5 meV due to the biexciton binding energy. Therefore, excited-state absorption pathways lead to red-shifted emission signals \cite{Miller:1982aa}.\\

\begin{figure}[t!]
\centering
  \includegraphics[width=0.8\textwidth]{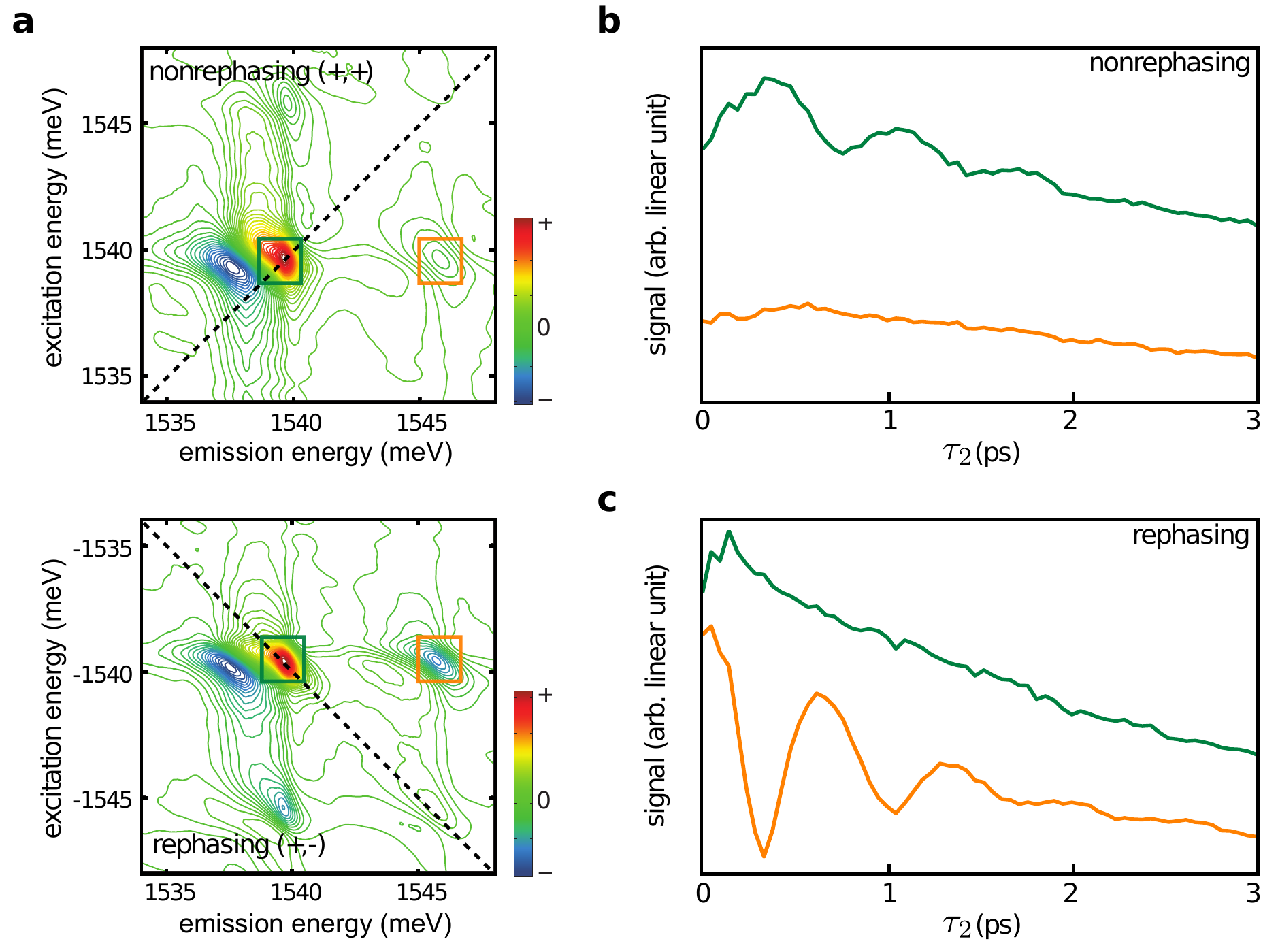}
  \caption{Model heterodimer system. 2D spectrum produced from a GaAs quantum well sample under cross-linear polarization configuration. Heavy-hole and light-hole excitons are excited at 1540 and 1546 meV, respectively. (a) The rephasing and nonrephasing measurements for $\tau_2 = 0$. (b) and (c)  Extractions from the heavy-hole exciton diagonal peak and one cross peak. Only the nonrephasing (rephasing) contribution to the diagonal (cross) peak oscillates. Data provided by D. B. Turner and K. A. Nelson (previously unpublished).}
  \label{fig:GaAs2d}
\end{figure}

The rephasing and nonrephasing measurements for $\tau_2 = 0$ are shown in Fig. \ref{fig:GaAs2d} (a) and (b), respectively. The bright diagonal peak is due to the heavy-hole exciton. We show extractions from two peaks for $0 \le \tau_2 \le 3$ ps in Fig. \ref{fig:GaAs2d}(c). The extractions match the predictions for the heterodimer in Fig. \ref{fig:2dspec}, where only the rephasing contribution to the cross peak oscillates and only the nonrephasing contribution to the diagonal peak oscillates \cite{Cheng:2008aa,Turner:2011ab,Turner:2012ab}. Excitation conditions followed refs \cite{Turner:2009aa,Turner:2010aa,Turner:2011aa,Turner:2012ac}.

\section{Double-sided Feynman diagrams}\label{sec:diagrams}

\begin{figure}[t!]
\centering
  \includegraphics[width=0.8\textwidth]{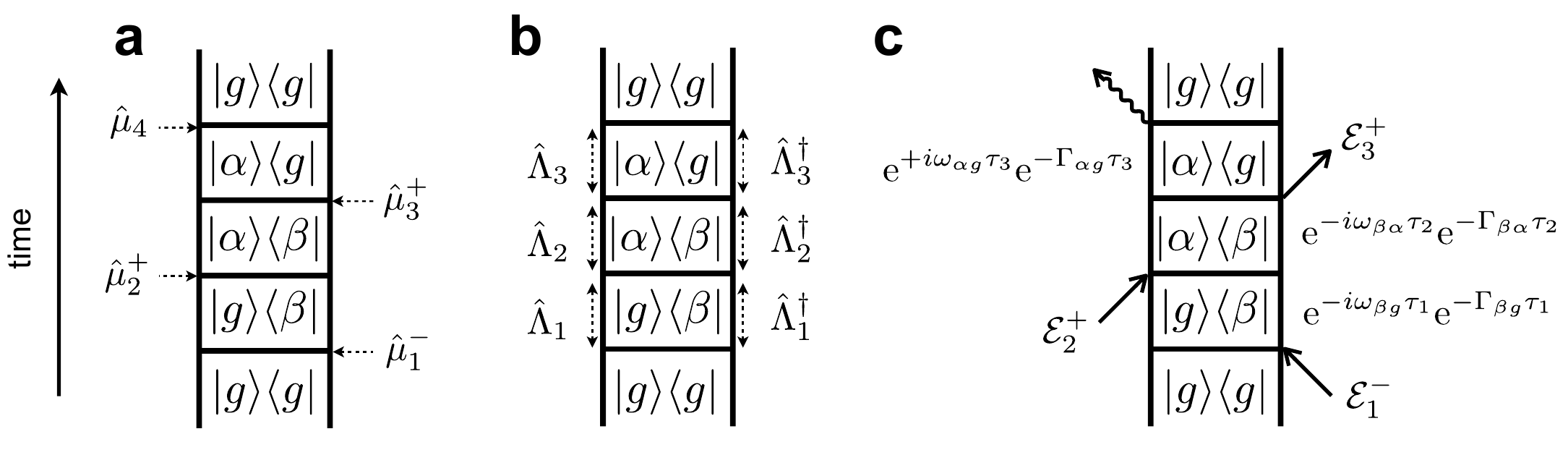}
  \caption{A double-sided Feynman diagram that follows the evolution of a single density matrix element (Eq. (\ref{eq:thing7})) as a function of time. (a) Horizontal lines represent the action of a transition dipole moment operator on the density matrix element. (b) Vertical lines represent the evolution of the density matrix element. (c) We can infer that whenever $\hat{\mu}_{j}^{\pm}$ acts on the density matrix element, we must include the relevant electric field term $\mathcal{E}^{\pm}_{j}$. The field contribution is typically depicted by a diagonal arrow, pointing to the right for $\mathcal{E}^+_{j}$ (containing a $+\mathbf{k}_{j}$ component) or pointing to the left for $\mathcal{E}^-_{j}$ (containing a $-\mathbf{k}_{j}$ component). The curvy arrow represents the emission of a single photon and the final density matrix element must return to a population term, i.e. $\ket{j}{}\bra{j}{}$, to be considered a valid diagram. The phase of each element oscillates at $\omega_{ab}=\omega_{a}-\omega_{b}$; $\Gamma_{aa}$ is the population relaxation rate for eigenstate $\ket{a}{}$, and $\Gamma_{ab}$ is the dephasing rate between eigenstates $\ket{a}{}$ and $\ket{b}{}$.  }
  \label{fig:DSFD}
\end{figure}

For complicated systems involving multiple chromophores, expansions such as Eq. (44) can involve a great number of terms.  Often only a few terms contain physically interesting information. Double-sided Feynman diagrams are a diagrammatic method for illustrating and evaluating individual terms in the expression for the generated signal. To demonstrate the relationship, consider the term in Eq. (\ref{eq:resp7d}) for a rephasing experiment, with the transition dipole moment operator defined in Eq. (\ref{eq:tdo}) (recall that for a rephasing experiment, the analytic signal is given by only the conjugate terms in Eq. (\ref{eq:resp})). Using the definition of the transition dipole moment operator in the interaction picture, given in Eq. (\ref{eq:mut}), we can rewrite the complex conjugate of the term in  Eq. (\ref{eq:resp7d}) as
\begin{align}\label{eq:thing4}
F_{2}^{-++}(\tau_{1},\tau_{2},\tau_{3})^*={}&-\mathcal{E}^-_{1}\mathcal{E}^+_{2}\mathcal{E}^+_{3}{}\mathrm{Tr}[\hat{\mu}_{4}\hat\Lambda_{3}\hat\Lambda_{2} \hat{\mu}_{2}^+\hat\Lambda_{1} \hat{\rho}_{0}\hat{\mu}_{1}^-\hat\Lambda\dg_{1}\hat\Lambda\dg_{2}\hat{\mu}_{3}^+\hat\Lambda\dg_{3} ]\,,
\end{align}
where $\mathcal{E}^+_{j}$ is defined in Eq. (\ref{eq:tildeE}) and
\begin{align}
\hat{\Lambda}_{j}={}&\ket{g}{}\bra{g}{}+e^{-i(\omega_{\alpha}-i\gamma{\alpha})\tau_{j}}\ket{\alpha}{}\bra{\alpha}{}+e^{-i(\omega_{\beta}-i\gamma{\beta})\tau_{j}}\ket{\beta}{}\bra{\beta}{}+e^{-i(\omega_{f}-i\gamma_{f})\tau_{j}}\ket{f}{}\bra{f}{}\,.
\end{align}
We recall that at this stage, $\gamma_{i}$ does not have a physical interpretation itself. Later, we make the substitution $(\gamma_{a}+\gamma_{b})\tau_{k}\rightarrow \Gamma_{ab}\tau_{k}$ and  recognize $\Gamma_{aa}=1/T_{1}$ as the population relaxation rate for eigenstate $\ket{a}{}$, and $\Gamma_{ab}=1/T_{2}$ as the dephasing rate between eigenstates $\ket{a}{}$ and $\ket{b}{}$. The positive frequency part of the transition dipole moment operator is given by
\begin{align}
\begin{split}
\hat{\mu}_{j}^+={}&\mu_{\alpha g}\ket{{\alpha}}{}\bra{{g}}{}+\mu_{\beta g}\ket{{\beta}}{}\bra{{g}}{}+\mu_{f\alpha}\ket{{f}}{}\bra{{\alpha}}{}+\mu_{f\beta}\ket{{f}}{}\bra{{\beta}}{}\,,
\end{split}
\end{align}
where the subscript $j$ is used to track the relationship between the positive and negative frequency components that results from the RWA, i.e. $\mathcal{E}^{\pm}_{j}\hat{\mu}_{j}^{\pm}$. Eq. (\ref{eq:thing4}) will contain quite a number of terms.  We consider just one:
\begin{align}\label{eq:thing7}
\begin{split}
\mathrm{term}={}&\mathcal{E}^-_{1}\mathcal{E}^+_{2}\mathcal{E}^+_{3}{}\mathrm{Tr}\Big[\underbrace{\mu_{\alpha g}\ket{{g}}{}\bra{{\alpha}}{}}_{\hat{\mu}_{4}}\underbrace{e^{-i(\omega_{\alpha}-i\gamma_{\alpha})\tau_3}\ket{\alpha}{}\bra{\alpha}{}}_{\hat\Lambda_{3}}\underbrace{e^{-i(\omega_{\alpha}-i\gamma_{\alpha})\tau_2}\ket{\alpha}{}\bra{\alpha}{}}_{\hat\Lambda_{2}}\underbrace{\mu_{\alpha g}\ket{{\alpha}}{}\bra{{g}}{}}_{\hat{\mu}^+_{2}}\underbrace{e^{-i(\omega_{g}-i\gamma_{g})\tau_1}\ket{g}{}\bra{g}{}}_{\hat\Lambda_{1}}\\
&\times\underbrace{\ket{g}{}\bra{g}{}}_{\hat{\rho}_{0}}\underbrace{\mu_{\beta g}\ket{{g}}{}\bra{{\beta}}{}}_{\hat{\mu}^-_{1}}\underbrace{e^{+i(\omega_{\beta}+i\gamma_{\beta})\tau_1}\ket{\beta}{}\bra{\beta}{}}_{\hat\Lambda\dg_{1}}\underbrace{e^{+i(\omega_{\beta}+i\gamma_{\beta})\tau_2}\ket{\beta}{}\bra{\beta}{}}_{\hat\Lambda\dg_{2}}\underbrace{\mu_{\beta g}\ket{{\beta}}{}\bra{{g}}{}}_{\hat{\mu}^+_{3}}\underbrace{e^{i(\omega_{g}+i\gamma_{g})\tau_3}\ket{g}{}\bra{g}{}}_{\hat\Lambda\dg_{3}}    \Big]\,.
\end{split}
\end{align}

The double-side Feynman diagram for this term is shown in Fig. \ref{fig:DSFD}. The diagram follows the evolution of a single density matrix element as a function of time. The frame which houses the time-evolving density matrix element consists of horizontal and vertical lines. Horizontal lines represent the action of a transition dipole moment operator on the density matrix element, as depicted in Fig. \ref{fig:DSFD} (a). Vertical lines represent the evolution of the density matrix element, as depicted in Fig. \ref{fig:DSFD} (b). Recall that by making the RWA, we fixed the relationship $\mathcal{E}^{\pm}_{j}\hat{\mu}_{j}^{\pm}$. We can therefore infer that whenever $\hat{\mu}_{j}^{\pm}$ acts on the density matrix element, we must include the relevant electric field term $\mathcal{E}^{\pm}_{j}$. The field contribution is typically depicted by a diagonal arrow, pointing to the right for $\mathcal{E}^+_{j}$ (containing a $+\mathbf{k}_{j}$ component) or pointing to the left for $\mathcal{E}^-_{j}$ (containing a $-\mathbf{k}_{j}$ component), as shown in Fig. \ref{fig:DSFD} (c). The curvy arrow represents the emission of a single photon and the final density matrix element must return to a population term, i.e. $\ket{j}{}\bra{j}{}$, to be considered a valid diagram.  A double-sided Feynman diagram contains all information required to infer the contribution to the emitted signal from that particular term. \\

Finally, although immensely valuable, double-sided Feynman diagrams have certain limitations.  For example, they cannot explain many of the most interesting phenomena that experiments have observed, many-body interactions, for example. Sometimes these phenomena, coherence transfer for instance, can be later drafted onto the double-sided Feynman diagrams.  Regardless, double-sided Feynman diagrams are almost always the first method one uses to describe measured signals.

\section{Concluding remarks}\label{sec:conc}
Two-dimensional electronic spectroscopy is a tremendously useful tool for characterizing a variety of systems such as atoms, molecules, molecular aggregates and related nanostructures, biological pigment-protein complexes and semiconductor nanostructures. Its utility arrises from the simultaneous and complex exploitation of multiple features such as phase-matching, pulse ordering and non-linear light-matter interactions. An unfortunate consequence of this is that the barrier to entry for researchers who would like to develop even a basic understanding of these techniques can sometimes be too high, and the interpretation of results is left to the experts.\\

For emerging interdisciplinary fields such as those revolving around coherence in energy transfer in photosynthesic systems---where a large portion of experimental studies are performed using nonlinear spectroscopic techniques  such as 2D ES---it is crucial for even non-experts in spectroscopy to understand the basics of how to interpret the experimental evidence.\\

We anticipate that our pedagogical guide will help such researchers, in particular those with a background in quantum optics, to understand the basics of 2D ES. Furthermore, we hope that this will expand the utility of this technique to areas not traditionally studied with 2D ES, such as Bose-Einstein condensates \cite{Pitaevskii2003}. 

\section{Acknowledgements}
We thank DARPA for funding under the QuBE program and the Natural Sciences and Engineering Research Council of Canada. DBT thanks Francesca Fassioli and Scott McClure for helpful discussions. AMB thanks Jessica Anna, Chanelle Jumper, Tihana Mirkovic, Daniel Oblinsky, Evgeny Ostroumov  and John Sipe for helpful discussions. We also thank John Sipe and Heinz-Peter Breuer for thoughtful comments on earlier versions of this manuscript. We are extremely grateful to Keith A. Nelson for kindly allowing us to use the GaAs data. 


\providecommand{\WileyBibTextsc}{}
\let\textsc\WileyBibTextsc
\providecommand{\othercit}{}
\providecommand{\jr}[1]{#1}
\providecommand{\etal}{~et~al.}

\end{document}